\documentclass[fleqn,usenatbib,useAMS]{mnras}
\usepackage{graphicx}	
\usepackage{amsmath}	
\usepackage{amssymb}	
\usepackage{multicol}   
\usepackage{bm}		    
\usepackage{pdflscape}	
\usepackage[T1]{fontenc}
\usepackage{ae,aecompl}
\usepackage{newtxtext,newtxmath}
\usepackage{booktabs}  

\newcommand{\teff}{$T_{\rm eff}$}

\newcommand{\metal}{\mbox{[Fe/H]}}

\newcommand{\rapo}{r_\ensuremath{\mathrm{apo}}}
\newcommand{\rperi}{r_\ensuremath{\mathrm{peri}}}
\newcommand{\zmax}{Z_\ensuremath{\mathrm{max}}}

\title[Sr and Ba signatures in metal-poor main-sequence stars]{{Metal-poor stars observed with the Magellan Telescope. IV. \\ Neutron-capture element signatures in 27 main-sequence stars}\thanks{This paper includes data gathered with the 6.5\,m Magellan Telescopes located at Las Campanas Observatory, Chile.}}
\author[Mohammad K.\ Mardini et al.]{
Mohammad K.\ Mardini,$^{1,2,3,4}$\thanks{E-mail: \href{mailto:m.mardini@ipmu.jp}{m.mardini@ipmu.jp}}
Anna Frebel,$^{3}$
Leyatt Betre,$^{3}$
Heather Jacobson,$^{3}$
John E. Norris,$^{5}$
\newauthor
~and Norbert Christlieb,$^{6}$
\\
$^{1}$Kavli IPMU (WPI), UTIAS, The University of Tokyo, Kashiwa, Chiba 277-8583, Japan\\
$^{2}$Institute for AI and Beyond, The University of Tokyo 7-3-1 Hongo, Bunkyo-ku, Tokyo 113-8655, Japan\\
$^{3}$Department of Physics and Kavli Institute for Astrophysics and Space Research, Massachusetts Institute of Technology, Cambridge, MA 02139, USA\\
$^{4}$Joint Institute for Nuclear Astrophysics–Center for Evolution of the Elements (JINA-CEE), East Lansing, MI 48824, USA\\
$^{5}$Research School of Astronomy \& Astrophysics, The Australian National University, Weston, ACT 2611, Australia\\
$^{6}$Zentrum f\"ur Astronomie der Universit\"at Heidelberg, Landessternwarte, K\"onigstuhl 12, 69117 Heidelberg, Germany
}

\begin{document}
\label{firstpage}
\pagerange{\pageref{firstpage}--\pageref{lastpage}}
\maketitle

\begin{abstract}
Based on high-resolution spectra obtained with Magellan/MIKE, we present a chemo-dynamical analysis for 27 near main-sequence turnoff metal-poor stars, including 20 stars analyzed for the first time. The sample spans a range in [Fe/H] from $-2.5$ to $-3.6$, with 44\% having \metal $<-2.9$. We derived chemical abundances for 17 elements, including strontium and barium. We derive Li abundances for the sample, which are in good agreement with the ``Spite Plateau'' value. A dozen of stars are carbon-enhanced, i.e., [C/Fe] $>0.7$. The lighter elements ($Z<30$) generally agree well with those of other low-metallicity halo stars. This broadly indicates chemically homogeneous gas  at the earliest times. Of the neutron-capture elements, we only detected strontium and barium. We used the [Sr/Ba] vs. [Ba/Fe] diagram to classify metal-poor stars into five populations based on their observed ratios. We find HE~0232$-$3755 to be a likely main $r$-process star, and HE~2214$-$6127 and HE~2332$-$3039 to be limited-$r$ stars. CS30302-145, HE~2045$-$5057, and CD~$-$24{\textdegree}17504 plausibly originated in long-disrupted early dwarf galaxies as evidenced by their [Sr/Ba] and [Ba/Fe] ratios. We also find that the derived [Sr/H] and [Ba/H] values for CD~$-$24{\textdegree}17504 are not inconsistent with the predicted yields of the $s$-process in massive rotating low-metallicity stars models. Further theoretical explorations will be helpful to better understand the earliest mechanisms and time scales of heavy element production for comparison with these and other observational abundance data. Finally, we investigate the orbital histories of our sample stars. Most display halo-like kinematics although three stars (CS~29504-018, HE~0223$-$2814, and HE~2133$-$0421) appear to be disk-like in nature. This confirms the extragalactic origin for CS~30302-145, HE~2045$-$5057, and, in particular, CD~$-$24{\textdegree}17504 which likely originated from a small accreted stellar system as one of the oldest stars.
\end{abstract}




\section{Introduction}

The chemical evolution of the Galaxy and the early Universe is a key topic in modern astrophysics. By studying the atmospheric composition of the oldest Galactic field metal-poor stars, we can obtain observational constraints on the nature of the nucleosynthetic yields of the first generation(s) of supernovae (SNe) (e.g., \citealt{UmedaNomoto:2002, Keller2014, kobayashi14}), early low-mass star formation processes \citep{dtrans, schneider12b, ji14} as well as cosmological models of the first stars and galaxies (e.g., \citealt{bromm02,greif08,bromm_araa11, wise12}), thereby enabling us to reconstruct the physical and chemical conditions of the earliest times. For example, the apparent uniformity in the observed light-element abundances provide important clues to firmly establish what was chemical ``normality'' at the earliest times; indicating a well-mixed medium or well-distributed population of similar SNe or both \citep[][]{Cayrel1996,cayrel2004}. Also, it provides an indication to what extent individual SNe patterns could have been preserved long enough to allow the formation of metal-poor stars from the ejecta without significant dilution or mixing \citep[e.g.,][]{frebel10, chan17,Mardini2022b}.

In contrast, light neutron-capture elements (e.g., Sr, Y, Zr) spoiled this uniformity by displaying a large abundance spread (reaching nearly 1.5\,dex) in comparison with the solar $r$-elements abundance pattern (see figures~5 and 11 in \citealt{Frebel2018} and \citealt{sneden_araa}). Also, \citet{aoki05} showed that stars with high [Sr/Ba] ratios are particularly evident at extremely low metallicity ($\metal<-3.0$)\footnote{\metal = $\log_{10}(N_{\text{Fe}}/N_{\text{H}})_{\star}-\log_{10}(N_{\text{Fe}}/N_{\text{H}})_{\sun}$}. In attempt to interpret these observations, \citet{aoki05} suggested the weak/limited $r$-process \citep[see also][]{wanajo01, wanajo05, izutani,Frebel2018}. The limited $r$-process would synthesize the necessary amounts of light neutron-capture elements, including Sr, with little (or no) heavy neutron-capture elements, such as Ba. The limited $r$-process has been theoretically studied and associated with massive core-collapse SNe, which is a requirement for the origin of these elements in the most metal-poor stars since type\,II SNe could have been the only SNe at these early times \citep[e.g.,][]{truran02,travaglio}. In addition, \citet{izutani} supported this hypothesis by illustrating that massive (25\,M$_{\odot}$) energetic hypernovae may produce large amounts of light neutron-capture elements in agreement with values observed in metal-poor stars with large [Sr/Ba] ratios.

Therefore, the [Sr/Ba] ratio is an important diagnostic tool to learn about the nucleosynthetic history that led to the abundance patterns observed in a given star. However, most available data for these elements are derived from giant stars, which might suffer from internal mixing. In contrast, metal-poor main-sequence stars have shallow convection zones and no internal mixing processes disturbing the stellar atmosphere, and thus these stars would provide the best-preserved chemical record of the early SNe. Considerably more observational time is required to obtain spectra useful for an accurate abundance analysis. Moreover, especially at very low metallicities, spectra with a much higher $S/N$ are required for main-sequence stars to ensure the detection of a sufficiently large number of atomic lines across the spectrum.

Among the first efforts to perform detailed abundance analysis of a large sample of metal-poor main-sequence stars are \citet{Carretta2002}, \citet{Cohen2004}, \citet{barklem05}. Idiosyncratically, the ``First Stars'' survey was designed to expand the chemical inventory for extremely metal-poor (EMP; \metal $<-3.0$) main-sequence stars \citep{cayrel2004}. In this work frame, \citet{bonifacio07} investigated Li abundances to study the level and constancy of the ``Spite Plateau'' at the
lowest metallicities. Also, \citet{bonifacio09} summarized the differences between the derived atmospheric abundances in various evolutionary statuses. Interestingly, JINAbase (database for metal-poor stars \citealt{jinabase}\footnote{Compilation available at~ \url{https://github.com/Mohammad-Mardini/JINAbase}}) shows that only 93 main-sequence stars\footnote{This number depends on the adopted temperature scales of the individual analyzes.} with \metal $<-3.0$ are known to have high-resolution spectroscopic abundance results available. Only $47\%$ of these 93 stars has a metallicity of \metal $<-3.3$. Therefore, further explorations of the [Sr/Ba] ratios observed in metal-poor main-sequence stars are desired to robustly characterize the $r$-limited process and gain insight into the (sophisticated) origin of these light neutron-capture elements.

In the era of the Gaia mission \citep[see][for more details about its scientific goals]{Gaia_the_mission}, a more precise picture of the formation and evolution of the Milky Way becomes feasible \citep[for recent review see][]{Helmi_review2020}. Linking the chemical abundances and kinematics of the most primitive ([Fe/H] $< -3.0$) stars provides direct insights on the early Galaxy, and by implication, of spiral galaxies in general \citep{Chiti_map}. With the improvements in astrometric measurements reported in the Gaia third data release \citep[hereafter Gaia DR3][]{Gaia_DR3} we can further investigate the formation and evolution of the individual Galactic components: thin disk \citep{Mardini2022b}, the Atari Disk \citep[or metal-weak thick disk e.g.,][]{Mardini2022}, thick disk and halo \citep[e.g.,][]{GSE_Belokurov2018,GSE_Helmi2018}. However, the bottleneck is to sophistically assign these stars to the various Galactic components. In \cite{Mardini2022}, we developed a sophisticated tool to robustly overcome the aforementioned problem. Employing this technique will not only allow us to investigate the origin of our sample but also to clean up the chemical signature based on the astrophysical formation site(s). Here we caution the reader that using a naive approach (e.g., the height from the disk) would yield some contamination from the other components.

This paper presents results from an ongoing observing program using the Magellan-Clay telescope and the MIKE spectrograph at Las Campanas Observatory. The aim is to observe stars with \metal $\lesssim-3.0$ to discover and characterize the population of metal-poor halo stars. It follows three earlier papers in this series, \citep{placco13}, \citet{placco2014a}, and \citet{placco15} on metal-poor giants with and without $s$-process enhancement. The observations of warm metal-poor stars (with no gravity information) also yielded a sample of metal-poor horizontal branch stars which will be presented in a forthcoming paper. The current paper is organized as follows. In Section~\ref{Sec:selection} we describe the selection of the target stars, the observations, and the analysis techniques. In Section~\ref{sec:abund_param}, information on the stellar parameter determination, abundance analysis, and uncertainties are given. We discuss the abundance results in Section~\ref{sec:discusion}. We investigate the observed the chemical signature of the sample with respect to potential nucleosynthetic origins in Section~\ref{Sec:discussion} and conclude in Section~\ref{Sec:summary}.

\section{Target selection and observations}\label{Sec:selection}

The target stars in our sample were first observed during several runs between 2007 and 2008 with the Australian National University's 2.3\,m telescope/Double Beam Spectrograph combination at the Siding Spring Observatory as part of the program that selects metal-poor candidates \citep{hes4} from the Hamburg/ESO objective-prism survey \citep{hespaperI}. The overall scientific goal of this long-term project was the discovery of metal-poor stars having $\metal < -2.0$ and the effort lead to the discovery of many new stars with $\metal < -3.5$ \citep[e.g.,][]{HE0107_Nature, HE1327_Nature, norris12_I,yong13a}. These medium-resolution spectra have a resolving power of $R~\sim~2000$ and cover the wavelength range 3600--5400\,\AA. The data were reduced with the package FIGARO \footnote{see \url{http://ascl.net/1203.013}}. Metallicity estimates were determined using the Ca\,K\,II line index together with the calibration presented in \citet{BeersCaKII}. All sample stars had \metal\, estimates of $\metal\lesssim -2.8$.

\begin{figure}
\includegraphics[width=0.49\textwidth]{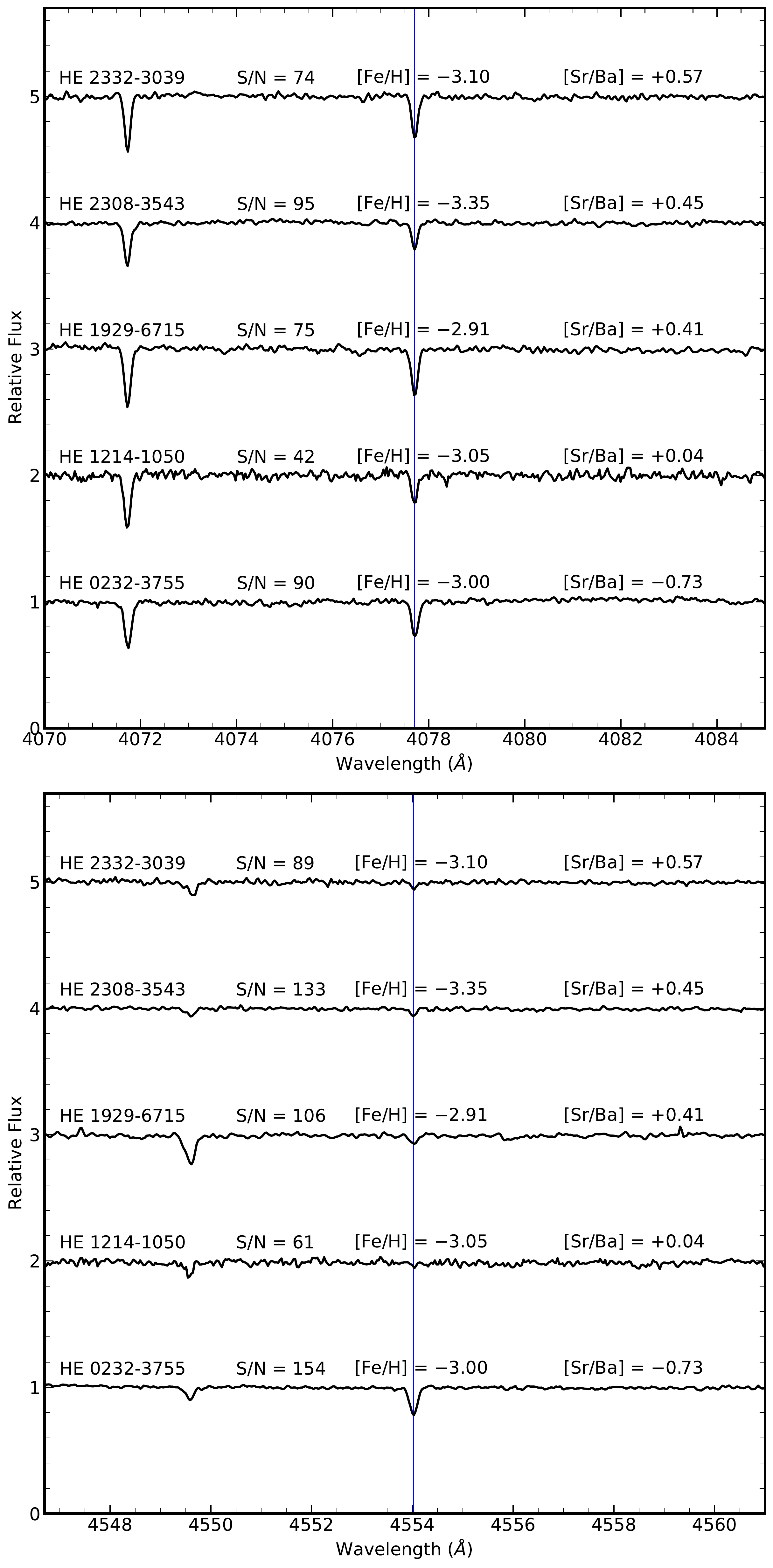}
\caption{Examples of the spectral region around the Sr\,II line at 4077\,\AA\ (upper) and the Ba\,II line at 4554\,\AA\ (lower), taken from five sample stars with metallicities ranging from $\metal=-2.91$ to $-3.35$. S/N ratios are measured around the feature of interest. [Fe/H] and [Sr/Ba] ratios are listed for illustration.}
\label{fig:specplot}
\end{figure}

During several runs in 2009, 2010, 2011, and 2014, our sample was then observed with the MIKE spectrograph on the Magellan-Clay telescope at Las Campanas Observatory \citep{mike} as part of a large high-resolution spectroscopy follow-up campaign of metal-poor star candidates with $\metal\lesssim-3.0$. The employed $0\farcs7$ slit width yields a resolving power of $R\sim35,000$ in the blue and $\sim~28,000$ in the red. In only two cases, the $1\farcs0$ slit was used which yields $R\sim28,000$ in the $\sim22,000$, respectively. CCD on-chip binning $2\times2$ was always applied. The wavelength coverage of the spectra is $\sim3400$-9000\,\AA. We observed the well-studied metal-poor star G64-12 (e.g., \citealt{1981G64-12, frebel13,g64_12_ref1,g64_12_ref2,g64_12_ref3}) as a comparison object and also included CD~$-$24{\textdegree}17504 \citep{Ryan1991,Jacobson2015}. Individual exposures, for our program stars, ranged from 5 to 90\,min to facilitate cosmic ray removal. The total exposure times ranged from $\sim10$\,min to 210\,min. See Table \ref{Tab:obs} for more details on the observations.

Reductions of the individual MIKE spectra were carried out using the MIKE Carnegie Python pipeline initially described by \citet{kelson03}\footnote{Available at \url{https://code.obs.carnegiescience.edu/mike}}. The wavelength calibration was accomplished using Th-Ar lamp frames. The reduced frames of the stars were normalized using a high-order cubic spline fits to the shape of each echelle order. The overlapping echelle orders were then merged into the final spectra. Two example $S/N$ ratio measurements of the final spectra are given in Table~\ref{Tab:obs}. Figure~\ref{fig:specplot} illustrates portions of several stars' spectra around the Sr\,II line at 4077\,\AA\ and the Ba\,II line at 4554\,\AA.

To measure the radial velocity ($v_{\rm{rad}}$) of each star, we used a cross-correlation technique to match the near-infrared Mg b triplet lines with those of HD140283 which we use as a metal-poor standard star template. For G64$-$12, we measure an average $v_{\rm{rad}} = 443.4$\,km\,s$^{-1}$, which is in good agreement with the well-established value of $v_{\rm{rad}} = 442.5$\,km\,s$^{-1}$ by \citet{lathamG64}. The measured RV value for CD~$-$24{\textdegree}17504 also in good agreement with \citet{CD_24_17504_RV}. Inspection of Table~\ref{Tab:obs} show that four stars have significant radial velocity variations: HE~0406$-$3120, HE~1214$-$1050, HE~1436$-$0654, and HE~2130$-$4852. One additional star, HE~1929$-$6715, shows variance at the $\sim$1\,km\,s$^{-1}$ level. Excluding the latter as a binary, the fraction of such stars in our sample is 4 of 27 (15\%).

\section{Abundance analysis}\label{sec:abund_param}
\subsection{Equivalent Width Measurements}

Equivalent width measurements are obtained by fitting Gaussian profiles to the lines. See Table~\ref{Tab:Eqw} for the lines used, their measured equivalent widths and line abundances. Given the low metallicity and warm temperature of our sample stars, only relatively few Fe and other elements lines are available across the entire wavelength range. For example, for iron -- having the most lines available of any element -- there are only $\sim50$ - $70$ Fe\,I lines available in the blue range, and less than 10 lines in the red spectra. For a few features (Li, CH, Sr, Ba) and upper limits, we use the spectrum synthesis approach. The abundance of a given species is obtained by matching the observed spectrum to a synthetic spectrum of known abundance. Elemental abundances were obtained using the latest version of the MOOG\footnote{Available at \url{https://www.as.utexas.edu/~chris/moog.html}} analysis code \citep{moog, sobeck11} together with 1D LTE Castelli \& Kurucz no-overshoot model atmospheres \citep{kurucz} and the SMHR software described in \citet{casey14}\footnote{Available at \url{https://github.com/andycasey/smhr}}.

\begin{figure*}
\begin{center}
\includegraphics[clip=true,width=\textwidth]{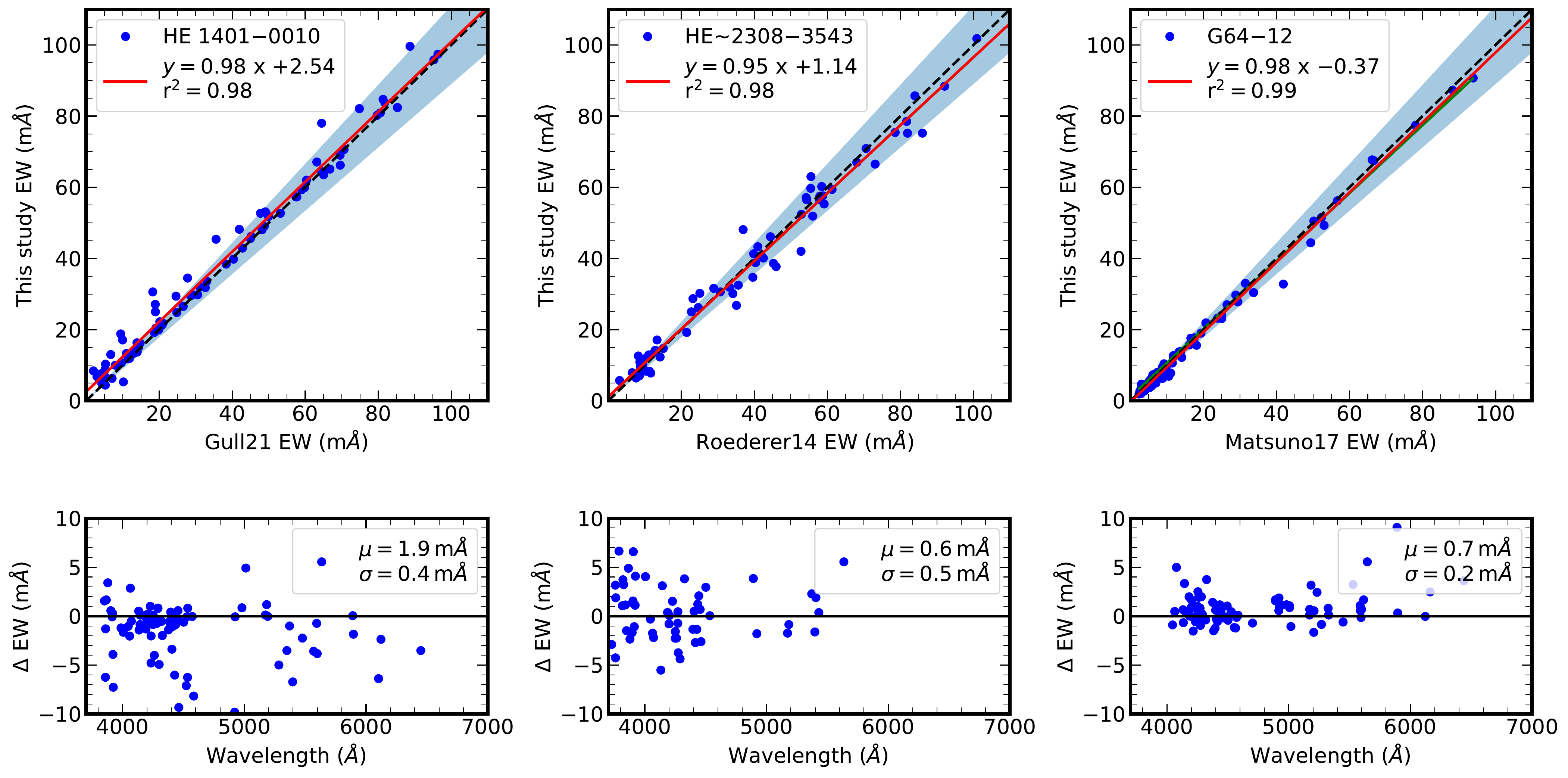}
\caption{Upper Panels show comparisons of the equivalent widths of HE~1401$-$001 (left), HE~2308$-$3543 (middle), and G64$-$12 (right), with \citet{Gull2021ews}, \citet{roederer14}, and \citet{Matsuno_g64ews}, respectively. The dashed lines represent one-to-one correlation. The shaded areas represent a 10\% tolerance. The solid lines represent best fit through the data. The solid line equations and the R-squared values are listed in the legends. Lower panels show the differences between equivalent widths' measurements ($\Delta$EW) as a function of wavelength. There is no obvious correlation between $\Delta$EW and wavelength. The mean of $\Delta$EW and standard error are listed in the legends\label{fig:EWs_comp}}
\end{center}
\end{figure*}

Figure~\ref{fig:EWs_comp} compares our equivalent width measurements with those reported in the literature for HE~1401$-$001 \citep{Gull2021ews}, HE~2308$-$3543 \citep{roederer14}, and G64$-$12 \citep{Matsuno_g64ews}. The upper panels show the one-to-one line (dashed line), the ``best fit'' (solid line) through the data, and 10\% tolerance (shaded areas). We also listed the solid line equations and the R-squared values in legends. The lower panels show the behavior of the equivalent widths discrepancies ($\Delta$EW) with wavelength. There is no obvious $\Delta$EW-wavelength correlation. We also listed the mean of $\Delta$EW and standard error ($\sigma$) in the legends. In general, the agreement is good, with mean offset $\mu = 1.9$\,m\AA\ (HE~1401$-$001), $\mu = 0.6$\,m\AA\ (HE~2308$-$3543), $\mu = 0.7$\,m\AA\ (G64$-$12). In summary, these $\Delta$EW values represent the varying quality of the spectra (see Table~\ref{Tab:obs}).

\subsection{Stellar Parameters}
We used the traditional spectroscopic method and the procedure outlined in \citet{frebel13} for stellar parameters determination. We constrain the effective temperatures by plotting the abundances of the measured Fe\,I lines as a function of their excitation potential (``excitation temperature''). We begin with rough initial temperature guesses according to the shape of the Balmer lines (after visually inspecting them against a set of stars with known temperatures). In general, spectroscopic techniques are known to yield cooler temperatures (and thus, lower surface gravities) than those obtained using photometry. This effect is most pronounced for cool giant stars but much less so for warmer stars near the main-sequence turnoff \cite[see][for more details]{frebel13}. Even though we are dealing with a sample of warm stars, we still apply temperature corrections according to $T_{\rm{eff,corr}} = T_{\rm{eff,initial}} - 0.1 \times T_{\rm{eff,initial}} + 670$, as given in \citet{frebel13}. We also employed the accurate and precise G, BP and RP magnitudes reported in \citet{Gaia_DR3} to estimate photometric effective temperatures, which are less affected by the inadequacies of the models of stellar atmospheres. We used the best polynomial fit reported by \citet{Mucciarelli2021}. We estimated the bolometric correction from \citet{Casagrande2018BC1,Casagrande2018BC2}. The extinction coefficients are calculated from the gird presented in \citet{Riello2021} \footnote{For Gaia DR3 are obtained using updated passband and Vega zero points from \url{https://www.cosmos.esa.int/web/gaia/edr3-passbands}}. The resulting $T_{\rm{eff,corr}}$ are in good agreement with those derived from photometry. The uncorrected, corrected spectroscopic, and photometric temperatures are listed in Table~\ref{Tab:stellpar}.

We determine the surface gravity from the ionization equilibrium of Fe\,I and Fe\,II abundances. In this iterative process, the microturbulence, \mbox{v$_{\rm micr}$}, is obtained by demanding no trend of individual Fe\,I line abundances with equivalent widths. The effective temperature uncertainties are estimated to be of the order of 100\,K (internal accuracy) and 200\,K regarding systematic differences compared with other temperature scales. We estimate the surface gravity uncertainties by searching for the range in gravities in which the Fe\,II abundance still matches the Fe\,I abundance within its uncertainties. For an Fe\,I uncertainty of 0.15\,dex this typically results in $\sigma\log (g)=0.5$\,dex. We adopt uncertainties in the micro turbulence of 0.3\,km\,s$^{-1}$ as indicated by the spread of individual line abundances. We also estimate $\log (g)_{\rm{Gaia\_plx}}$ using corrected parallaxes from the third data release of Gaia \citep{Gaia_DR3} and equations~2 and 3 from \citet{Mardini_2019b}. The resulting $\log (g)_{\rm{corr}}$ are in better agreement with those derived from distances. All stellar parameters are summarized in Table~\ref{Tab:stellpar}. We adopted the corrected stellar parameters for the remaining analysis. In summary, our sample stars span a metallicity range of $-3.61<\metal<-2.26$.

\begin{figure}
\begin{center}
\includegraphics[clip=true,width=8.5cm]{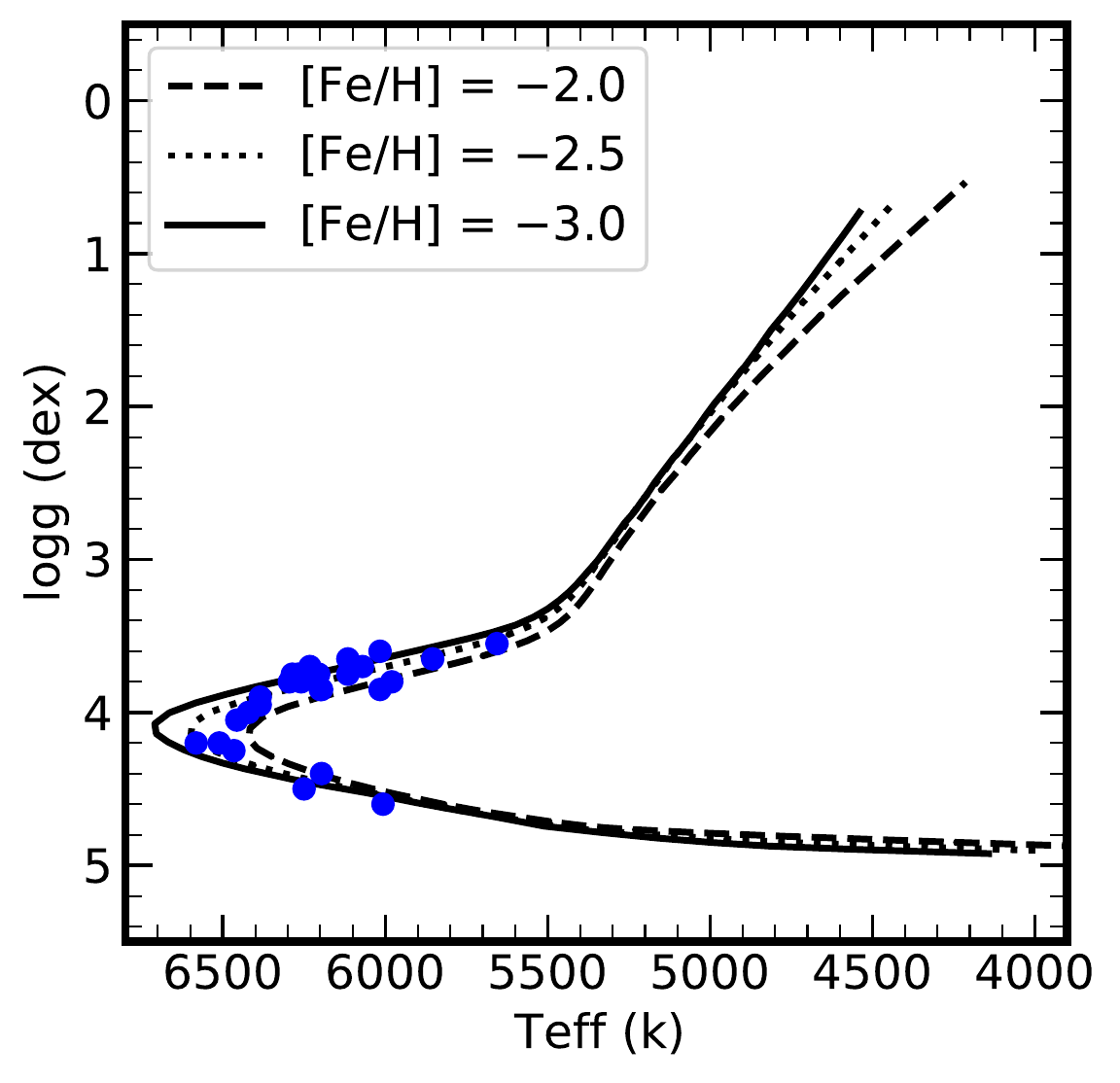}
\caption{Adopted surface gravity ($\log (g)_{\rm{corr}}$) as a function of effective temperature ($T_{\rm{eff,corr}}$) of our sample stars in comparison with 12\,Gyr isochrones with metallicities ranging from $\metal=-2.0$ to $-3.0$.\label{fig:iso}}
\end{center}
\end{figure}

For inspection of the robustness of the stellar parameters, we overplotted 12\,Gyr Dartmouth isochrone with an $\alpha$-enhancement of $\mbox{[$\alpha$/Fe]}=0.4$ and $\metal=-2.0, -2.5,$ and $-3.0$ \citep{dotter08_Dartmouth_iso}\footnote{Available at \url{http://stellar.dartmouth.edu/models/index.html}}. We find that our adopted stellar parameters are in good agreement with the isochrones. This is shown in Figure~\ref{fig:iso}. As can be seen, our sample relatively evenly covers the upper end of the main-sequence, the turnoff region and the subgiant branch. The derived abundances of our sample stars are presented in Table~\ref{Tab:abundances}. We comment on the individual elemental abundances and their nucleosynthetic origins in Section~\ref{sec:discusion}.

\subsection{Abundance Uncertainties}
In Table~\ref{Tab:abundances}, we list random uncertainties ($\sigma$), reflecting the standard deviations of individual line measurements for each element. For abundances measured from only one line, we adopt a nominal uncertainty of 0.15\,dex. For abundances measured from 2 or 3 lines but with unrealistically small uncertainties, we adopt a nominal value of 0.10\,dex. We test the robustness of our derived abundances by changing one stellar parameter at a time by its uncertainty. Table~\ref{Tab:err} shows the results for all elements, including the random uncertainties. Adding all these error sources in quadrature, the abundances have an overall uncertainty of $\sim0.15$ to 0.20\,dex.  Systematic uncertainties, e.g., arising from the choice of model atmosphere, are not a significant source of error. Uncertainties in $gf$ values have not been considered, but are expected to be of minor influence compared with the other uncertainties.

We also compare our stellar parameters and abundance results with those previously obtained for G64$-$12 \citep{frebel13} and CD~$-$24{\textdegree}17504 \citep{Jacobson2015}. Regarding the stellar parameters, temperatures are within 10\,K for both stars, although it should be noted that the same spectrum was used in the case of G64$-$12. A difference of 0.01\,dex in \metal\, was found for G64$-$12, but 0.20\,dex for CD~$-$24{\textdegree}17504. This difference is further described in \citet{Jacobson2015} and mostly due to the much higher S/N of their spectrum that enables measurement of additional Fe\,II lines.

\section{Chemical abundance signatures of 27 near Main Sequence turnoff stars}\label{sec:discusion}


\subsection{Lithium}

\begin{figure}
\begin{center}
\includegraphics[clip=true,width=8.5cm]{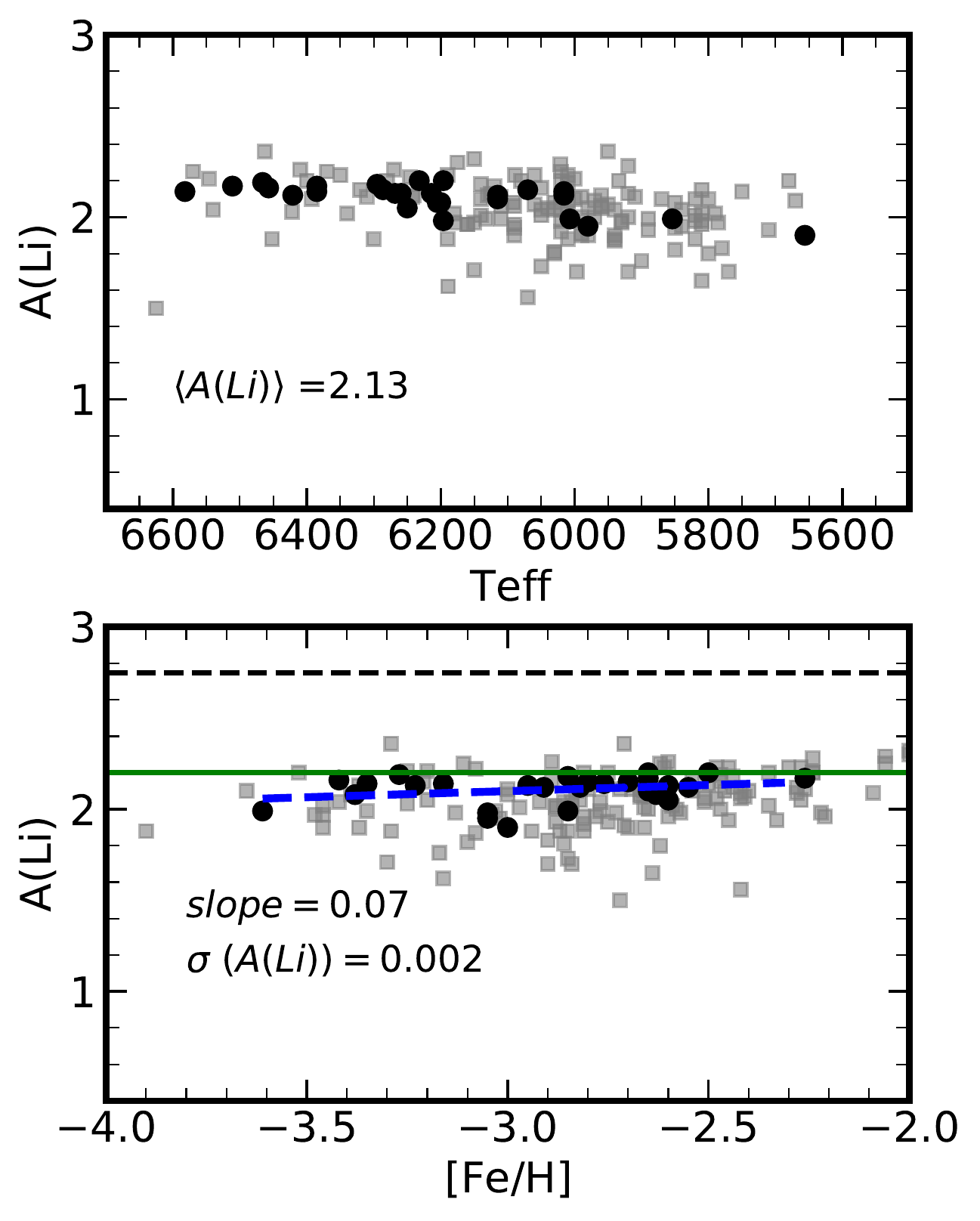} 
\caption{Lithium abundances $A(\rm Li)$ for our sample of stars (black points); as a function of effective temperature (top panel) and metallicity [Fe/H] (bottom panel). Gray squares represent dwarf stars ($5500 < T_{\rm{eff}} < 6700$, $\log (g) >3.5$, and \metal $<-1.5$) selected from JINAbase. The solid green line represents the ``Spite Plateau'' $A(\rm Li)$ = 2.23. Dashed black lines represent the primordial lithium $A(\rm Li)$ = 2.72. Dashed blue line represents the ``best-fit'' line through the data. The slope and residual are listed. Our sample has a mean lithium abundance of $A(\rm Li)$ = 2.13. }
\label{fig:li}
\end{center}
\end{figure}

As stars evolve from the main sequence towards the red giant branch, their surface convection zone deepens. The deeper layers towards the bottom of the convection zone are hot enough to destroy the fragile element lithium \citep{Lithium_depletion}. Typical main-sequence Galactic stars have Li abundances of $A(\rm {^7}Li)\sim2.2$ \citep{spite_lithium_pl82}. Given the evolutionary status of our sample, we detected the Li\,I doublet at 6707\,{\AA} in all of our spectra. We considered the Hyperfine splitting (HFS) effect and multiple isotopes in the syntheses of Li. The derived abundances are listed in Table~\ref{Tab:abundances}. In Figure~\ref{fig:li}, we show $A(\rm {^7}Li)$ abundances as a function of effective temperature (upper panel) as well as [Fe/H] (lower panel). We also selected similar dwarf stars ($5500 < T_{\rm{eff}} < 6700$, $\log (g) >3.5$, and \metal $<-2.0$) from JINAbase for comparison. We also plot the primordial lithium $A(\rm {^7}Li)$ = 2.75 (dashed black line) and our linear ``best-fit'' line (dashed blue line).

In general, our abundances are in good agreement with those found in the other dwarfs at the ``Spite Plateau'' (green solid line) although at a somewhat lower mean value of $A(\rm Li)$ = 2.13 (from stars with $6050 < T_{\rm{eff}} < 6600$). \citet{Norris2023_Li} extensively investigated the behavior of $A(\rm Li)$ as a function of \teff, [Fe/H], and [C/Fe]. Reanalyzing literature data for warm metal-poor dwarfs, they reconfirmed the ``meltdown'' of the Spite plateau \citep{bonifacio07} in the metallicity range of $-4.2<$ [Fe/H] $<-3.0$. While we find relatively flat trend for our warmest stars, our most metal-poor star (HE~2308$-$3543 with \teff = 6007\,K) has $A(\rm Li)$ = 2.0 which adds to stars with sub-Spite Plateau abundances.




%

\begin{figure*}                        
\begin{center}
\includegraphics[clip=true,width=\textwidth]{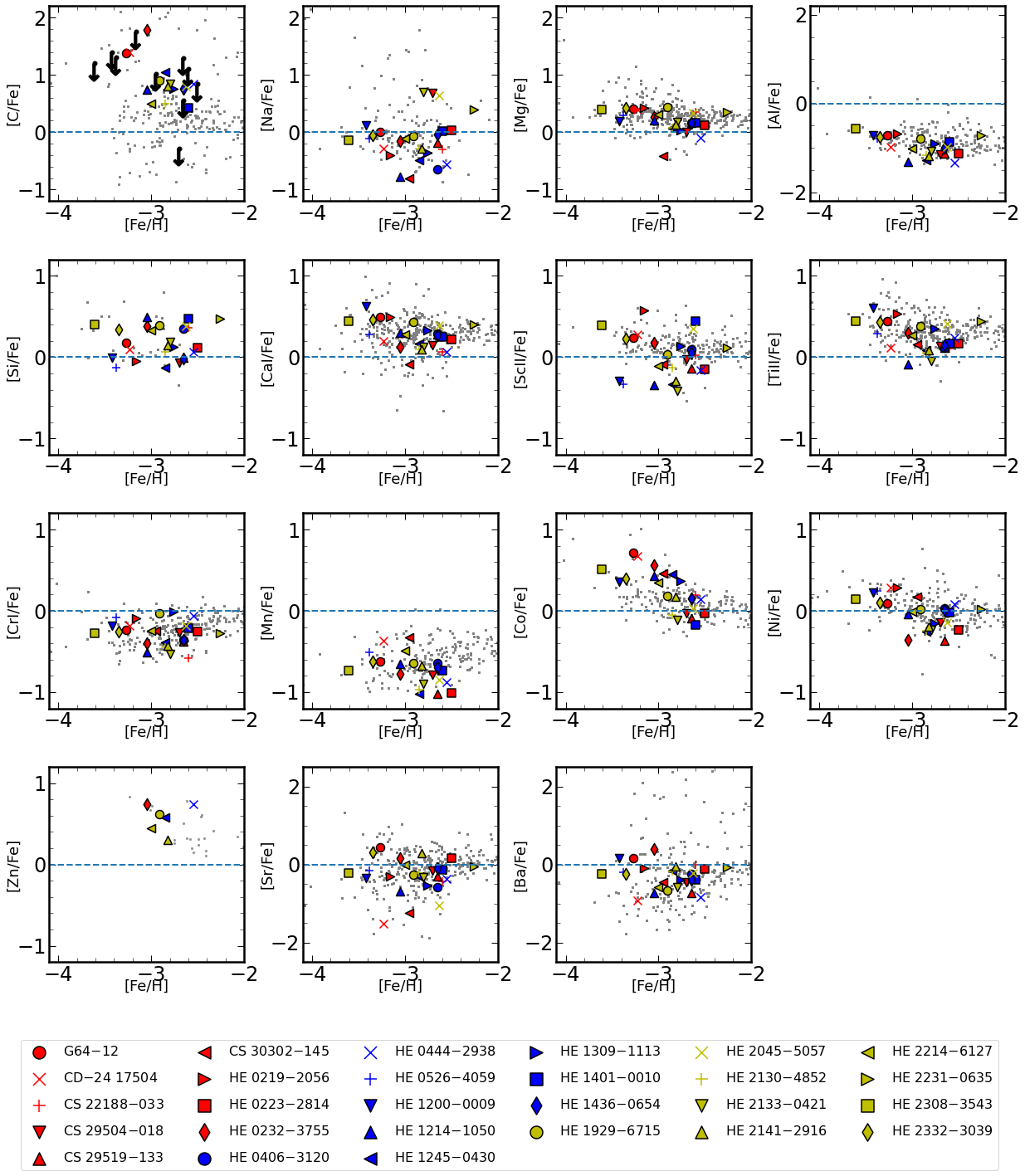}
\caption{Abundance ratios as a function of metallicity of our metal-poor stars (various symbols) and literature stars collected by JINAbase \citep{jinabase}. Data taken from: \citet{cayrel2004}; \citet{roederer14c}; \citet{barklem05}; \citet{Yong2013}; \citet{Aoki2013}; \citet{Jacobson2015}; \citet{Mardini_2019a,Mardini_2019b,Mardini_2020}.}
\label{fig:chemabunds}
\end{center}
\end{figure*}

\subsection{Light Elements: Carbon through zinc}
Figure~\ref{fig:chemabunds} shows the abundance results [X/Fe] as a function of metallicity [Fe/H] in comparison with other very metal-poor (\metal~$\lesssim -2.0$ ) stars collected by JINAbase from the literature \citep{jinabase}. Due to the low metallicity and the warm effective temperature of all sample stars, carbon measurements were only possible in 15 of 27 stars. For these 15 stars, when possible, we used the molecular G-band head at $4313$\,{\AA} for deriving $A(\rm C)$. In a few cases, the feature at 4323\,{\AA} was also used. For stars with no C detection, we derived upper limits which are rather high (between $\mbox{[C/Fe]}\sim0.5$ and 1.5) and unfortunately not very meaningful. 12 of the 15 stars with C detection have $\mbox{[C/Fe]}>0.7$ while four have $\mbox{[C/Fe]}>1.0$. Furthermore, there is a tendency for the most metal-poor stars to have the higher [C/Fe] values, as can be seen in Figure~\ref{fig:chemabunds}. Assuming that all stars with upper limits are not carbon-enhanced, the fraction of carbon-enhanced metal-poor stars in this sample would be 44\%. This number would only increase if any other stars with upper limits would also be carbon-enhanced. A fraction of $\sim44\%$, taking into account the \metal\ distribution of the sample, is in good agreement with the results of \citet{placco14b}. They established the fraction of carbon-enhanced metal-poor stars in the halo; 24\% for stars with $\metal\sim-2.0$ and 43\% for stars with $\metal\sim-3.3$ (see their figure~9).

We measure a suite of light-, $\alpha$- and Fe-peak elements: Na, Mg, Al, Si, Ca, Sc, Ti, Cr, Mn, Co, Fe, and Ni. Detailed comments on the elements and employed lines can be found in \citet{frebel_he1300} and will not be repeated here. In Figure~\ref{fig:chemabunds}, we compare our results with those of other metal-poor stars collected by JINAbase. Regarding these light elements ($Z<26$), all stars share very similar abundance ratios, [X/Fe], and appear to be typical halo stars. Only CS 30302$-$145 appears to have low $\mbox{[Mg,Ca/Fe]}$ ratios, HE~1200$-$0009 and HE~0526$-$4059 have high low $\mbox{[Sc/Fe]}$ ratio, and HE~1214$-$1050 has relatively low $\mbox{[Al,Sc,Ti/Fe]}$.

\citet{cayrel2004} noted that, especially at the lowest metallicities (in the range $-4.0\lesssim\metal\lesssim-3.0$), there is little scatter in the data indicating a chemically homogeneous medium already at the earliest times. The halo-typical abundances in our ten main-sequence stars at $\metal\sim-3.0$ are in line with this finding and furthermore support that there was a population of extremely metal-poor stars that formed from a medium enriched in light elements originating from the same process(es) and in similar quantities and elemental distributions. There are, however, also other stars (e.g., \citealt{McWilliametal,aoki_mg}) that significantly deviate from the general halo trends as found among the lighter elements. Below $\metal\sim-3.5$ the fraction of these ``outliers'' is around 20\% to 30\%, but none of our stars shows such behavior.

\subsection{Neutron-capture elements}

\begin{figure*}
\begin{center}
\includegraphics[clip=true,width=\textwidth]{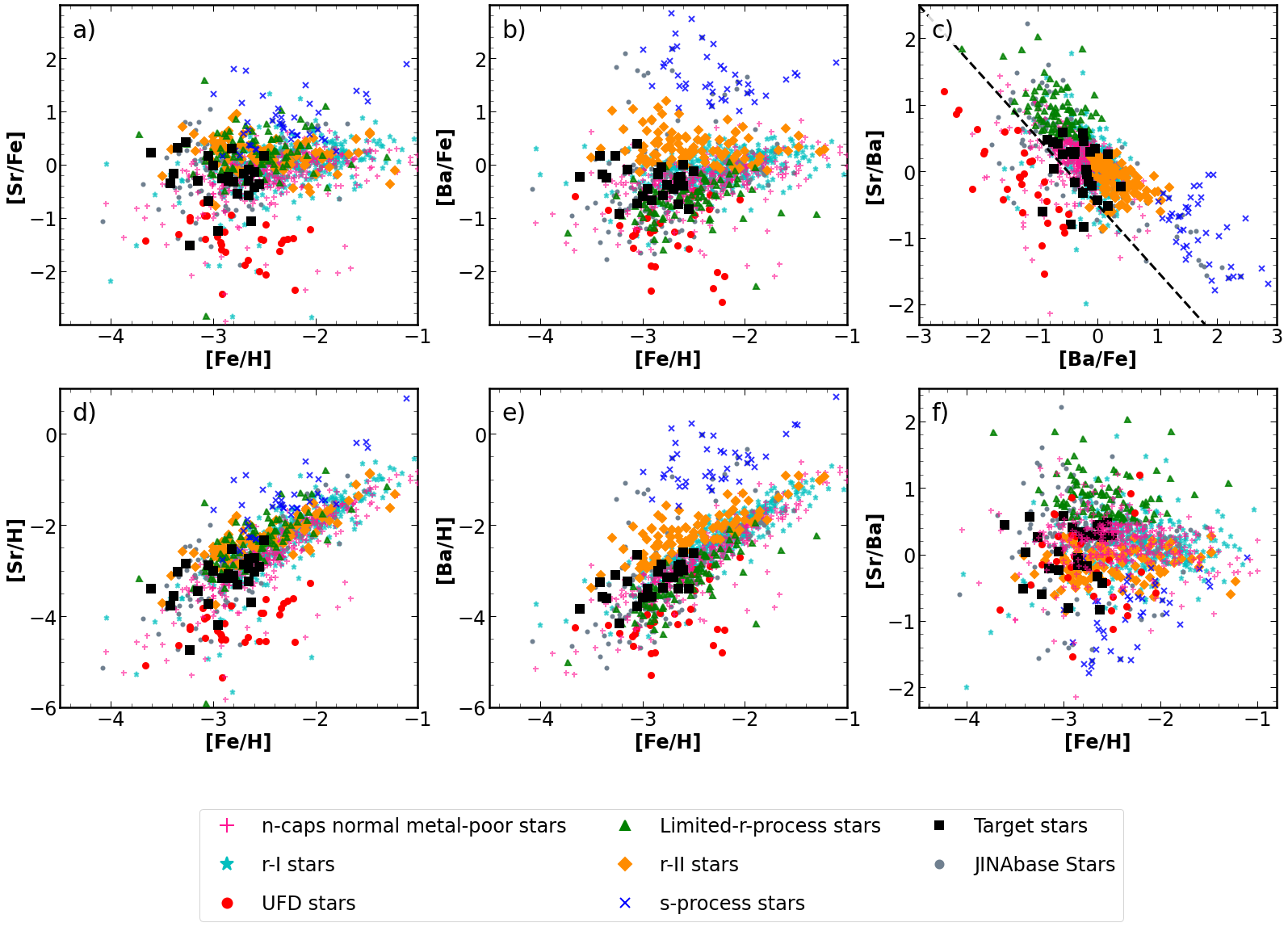}
\caption{Sr and Ba abundances for our sample stars in comparison with literature metal-poor stars. Cyan asterisks represent {r}-I stars. Green triangles represent limited $r$-process stars. Pink plus-symbols represent the neutron-capture normal metal-poor stars. Black squares represent our sample stars. Red circles represent ultra-faint dwarf galaxy stars. Orange diamonds represent main $r$-process stars. Blue crosses represent $s$-process stars. As can be seen there is a separation of these populations in the [Sr/Ba] vs. [Ba/Fe] space which helps to distinguish the origins of the chemical signatures of metal-poor stars.  
Metal-poor stars data taken from: 
\citet{roederer14c}; \citet{barklem05}; \citet{Ryan1991}; \citet{Cohen2004}; \citet{Hollek2011}; \citet{Bonifacio2012}; \citet{Mardini_2019a}; \citet{Aguado2017b}; \citet{Mardini_2019b}; \citet{Yong2013}; \citet{Frebel2010}; \citet{Hansen2015}; \citet{Mardini_2020}; \citet{Jacobson2015};
\citet{ezzeddine19};
\citet{Spite2014}; \citet{cayrel2004}; \citet{Norris2007}; \citet{Mardini2019c}; \citet{Aguado2021}; \citet{Lai2004}; \citet{Almusleh2021}; \citet{Placco2020}; \citet{Ryan1996}; \citet{Aguado2018}; \citet{Taani2019,Taani2019b}; \citet{Aoki2007}; \citet{Aoki2013}; \citet{Cohen2013}; \citet{Honda2011}; \citet{Lai2008}; \citet{Casey2015}; \citet{Li2015a}; \citet{Masseron2006}; \citet{Rich2009}; \citet{Taani_2022}; \citet{Ryan1999}; \citet{Spite1999}; \citet{Depagne2000}; \citet{Spite2000}; \citet{Sivarani2006}; \citet{Norris2001}; \citet{placco2014a}; \citet{Caffau2011a}; \citet{frebeletal05}; \citet{frebel08}; \citet{Li2015}; \citet{Behara2010}; \citet{Caffau2011}; \citet{Susmitha2016}; \citet{Aguado2017}; \citet{Carretta2002}; \citet{For2010}; \citet{2015PASJ...67...84L}; \citet{Keller2014}; \citet{frebel15b}; \citet{Caffau2013}; \citet{Plez2005}; \citet{Christlieb2002}; \citet{Frebel2019}. UFDs stars data taken from: \citet{Koch08}; \citet{Simon10}; \citet{frebel10}; \citet{Norris10b}; \citet{ji16a}; \citet{Koch13}; \citet{Gilmore13}; \citet{frebel14}; \citet{ji16b}; \citet{frebel16}; \citet{kirby2017_tri}; \citet{nagasawa2018}; \citet{Ji19}; \citet{Marshall19}; \citet{Ji20}; \citet{hansen20}; \citet{Waller22}; \citet{Chiti2023}}
\label{fig:srba}
\end{center}
\end{figure*}

Very few neutron-capture elements are measured for turnoff stars because their spectral lines are generally very weak. Unsurprisingly, we only detect two Sr\,II lines at $\lambda$4077 and $\lambda$4215 and two Ba\,II lines at 4554\,{\AA} and 4934\,{\AA} in our stars (see Figure~\ref{fig:specplot} for some examples). No other lines for neutron-capture elements were detected. HFS effects were taken into account when determining abundances from the Ba lines. The measured [Sr/Fe] ratio in our target stars ranges from $-$1.52 to +0.43\,dex. The [Ba/Fe] ratio ranges from $-$0.92 to +0.69\,dex. Figure~\ref{fig:chemabunds} shows principal agreement between the measured ratios ([Sr/Fe] and [Ba/Fe]) for our sample stars and those of literature stars, as collected by JINAbase. However, the overall abundance uniformity among the lighter elements does not hold for the neutron-capture elements which display a rather large star-to-star scatter, particularly at the lowest metallicities \citep[e.g.,][]{Frebel2018}.

In the literature, barium (Z = 56 mainly produced by the $s$-process) and europium (Z = 63 mainly produced by the $r$-process; see table~3 in \citealt{Prantzos2020}) are usually used as a quick tool to identify the underlying production mechanisms of the observed neutron-capture elements \citep[see table~1 in][]{Frebel2018}. Unfortunately, we could not derive any $A(\rm Eu)$ for our sample. Therefore, we could not use these classical abundance criteria to gain putative insight into the chemical enrichment of our stars. However, in Section~\ref{sec:mar-fre}, we show that the location of a star in the [Sr/Ba]-[Ba/Fe] space can be also used to comment on its possible chemical enrichment.

\section{Discussion}\label{Sec:discussion}

\subsection{Exploring the origins of neutron-capture elements}\label{sec:mar-fre}

The synthesis of the majority of elements with Z $>30$ is possible only through neutron-capture reactions when neutrons overcome the Coulomb barrier. The two neutron-capture ($s$- and $r$-) processes can be separated based on the timescales of the neutron-capture and associated $\beta$-decay. The $s$-process denotes that the neutron-capture is slower than the timescale of the $\beta$-decay of the nucleus involved. Nuclei along the so-called “valley of $\beta$-stability” are synthesized this way. In comparison, the $r$-process happens in more violent events and thus rapidly synthesize large amounts of unstable/radioactive nuclei far from stability \citep[e.g.,][]{Meyer1994}.

We now explore the origins of the observed neutron-capture elements by means of the [Sr/Ba] vs [Ba/Fe] behavior in our stars to better understand the formation environment in which these nuclei formed. In Figure~\ref{fig:srba}, we compare our results ([Sr/H,Fe]; left panels, [Ba/H,Fe]; middle panels, and [Sr/Ba]; right panels) as a function of [Fe/H] with the body of data of Galactic metal-poor stars collected by the JINAbase (small gray circles). We also selected all JINAbase stars which have Sr, Ba, and Eu measurements. The resulting sample contains 1069 distinct stars (shown with various symbols). Black squares represent our sample stars.  

In Figure~\ref{fig:srba}, we plot all stars collected by JINAbase on the background (small gray circles) and then overplot stars with Sr, Ba, and Eu measurements. For the latter ones, we use the classes and signatures of metal-poor stars presented in Table~1 in \citet{Frebel2018} to label their possible chemical signature. We begin by selecting orange diamonds (100 data points) which represent the strongly enhanced $r$-process stars with [Eu/Fe] $> +0.7$ and [Ba/Eu] $< 0.0$. Cyan asterisks represents mildly enhanced $r$-process stars (397 data points; $+0.3 <$ [Eu/Fe] $< +0.7$ and [Ba/Eu] $< 0.0$). Blue crosses represent $s$-process rich stars (40 data points; [Ba/Fe] $> +1.0$ and [Ba/Eu] $> +0.5$). Green triangles represent limited $r$-process stars (92 data points; [Eu/Fe] $< 0.3$, [Sr/Ba] $> 0.5$, and [Sr/Eu] $> 0.0$). Pink plus-symbols represents the remaining metal-poor (440 data points; neutron-capture normal) stars after subtracting out all the aforementioned signatures and classes. We also add ultra-faint dwarf (UFD) galaxy stars (red filled circles)\footnote{Compilation available at~ \url{https://github.com/alexji/alexmods/blob/master/alexmods/data/abundance_tables/dwarf_lit_all.tab}}.

Panels a and b in Figure~\ref{fig:srba} show the observed [Sr/Fe] and [Ba/Fe] as a function of [Fe/H]. No clear trend for metallicity can be observed. However, the most metal-poor ([Fe/H] $<-3.0$) stars seems to have the lower [Sr/Fe] and [Ba/Fe] values compared to the more metal-rich ones. This could be a direct sign for an early chemical enrichment. The production of the heavy elements occurred later in the Galaxy, because the $s$- and $r$-processes need seed nuclei (e.g., iron). Panels d and e in Figure~\ref{fig:srba} reveal a similar early chemical enrichment signature but a clear correlations between [Sr, Ba/H] and [Fe/H] are seen.

Panel f in Figure~\ref{fig:srba} shows [Sr/Ba] as a function of [Fe/H]. Interestingly, the [Sr/Ba] values, for all stars, are enveloped in the range of $-2.5<$ [Sr/Ba] $<2.5$. The upper limit of this range reflect $-1.5$ $-$~[Fe/H] of the lower limit of the [Fe/H] (i.e., $-4$). The lower limit of this range reflect $-1.5$ $+$~[Fe/H] of the upper limit of the [Fe/H] (i.e., $-1$). At lower metallicity, obvious scatter in the [Sr/Ba] values is clearly seen. This scatter might reflect the production of Ba by the $r$-process.

The dashed line in panel c in Figure~\ref{fig:srba} denotes [Sr/Ba] = 0.50 $-$ [Ba/Fe], which separates the regions populated by UFD stars from other Galactic metal-poor stars. Excluding Reticulum\,II stars \citep{ji16b}, the [Sr/Ba] ratio of the stars in the UFD galaxies region strongly depends on their [Ba/Fe] ratio and could be a signature of the earliest star-forming clouds \citep{Frebel_Norris_2015}. Regarding the [Sr/Ba] ratio for Galactic metal-poor stars, we find four features i) the main $s$-process stars are wildly dispersed ($\sim$ 2\,dex) for a given [Ba/Fe] and have the lowest [Sr/Ba] values. ii) the limited $r$-process stars show significant scatter ($\sim$ 1\,dex) in [Sr/Ba] and have the highest [Sr/Ba] values. iii) the strong $r$-process stars also shows significant scatter in [Sr/Ba] and bridge the gap between the $s$-process and limited $r$-process stars. iv) the mild $r$-process stars and neutron-capture normal stars significantly overlap and are centred at [Sr/Ba] = 0.0 and [Ba/Fe] = 0.0. No apparent separation can be easily found. 

Based on the bulk of each population, we suggest new abundance criteria to differentiate between the origins of neutron-capture enhancements. The observed neutron-capture elements in metal-poor stars with [Sr/Ba] $> 0.6$ and [Ba/Fe] $<0.0$ are most likely synthesized by the limited $r$-process. The neutron-capture elements observed in stars with [Sr/Ba] $<-0.6$ and [Ba/Fe] $>1.5$ were formed in a weak neutron flux environment associated with the main $s$-process. Stars with $-0.4 <$ [Sr/Ba] $<0.4$ and $0.0 <$ [Ba/Fe] $<1.0$ might have chemical signatures associated with the population of $r$-II stars. Stars with $-0.2 <$ [Sr/Ba] $<0.6$ and [Ba/Fe] $< 0.0 $ are most likely mild $r$-process star or are "neutron-capture normal", i.e. show no enhancement in neutron-capture elements.

Using the aforementioned abundance criteria, we find no stars with strong (external) $s$-process enrichment in our sample. Instead, the vast majority of our sample stars are most likely $r$-I and/or neutron-capture normal stars given their position in the [Sr/Ba] vs [Ba/Fe] diagram, i.e. they are mildy enhanced, or not enhanced in neutron-capture elements beyond the solar ratio. The exceptions are 1) HE~0232$-$3755 which is most likely an $r$-II star given its [Sr/Ba] ratio; 2) HE~2214$-$6127 and HE~2332$-$3039 which are most likely limited-$r$ stars (further discussed below); and 3) CS~30302-145, HE~2045$-$5057, and CD~$-$24{\textdegree}17504 which reside below the main trend of [Sr/Ba] vs. [Ba/Fe] as set by Galactic metal-poor stars (dashed line), in the region that is characteristically populated by UFDs stars, see Figure~\ref{fig:srba}. They may likely have a separate origin as further discussed below. 

We note that regardless of their neutron-capture abundance levels, while the production mechanisms (and sites) of lighter elements ($Z<30$) and (any) neutron-capture elements in SNe would be decoupled, we here assume that if a SN that produced any neutron-capture elements it would also generate lighter elements in somewhat ``standard'' quantities \citep{woosley_weaver_1995}. As such, a massive progenitor exploding as a SN would also provide $\alpha$-element abundances at the typical +0.3 to 0.4\,dex level. Since the individual Sr and Ba abundances, and also \metal, are relatively low for most of our sample stars, it appears not unlikely that our sample stars formed from gas chemically enriched by a single supernova or explosive event only. Indeed, as been shown by \citet{UmedaNomoto:2002} and subsequent works, it is possible to reproduce the observations of extremely metal-poor stars (e.g. \citealt{cayrel2004}) with the yields from just one massive energetic Pop\,III SN (``hypernova''), even when no elements with $Z>30$ are considered. In the following, we discuss specific scenarios that may help explain the origins of the neutron-capture elements in all of our sample stars.

\subsection{Neutron-capture nucleosynthesis processes}\label{sec:ncaps_origin}

In the subsequent sections, we explore in more detail possible chemical enrichment scenarios for the [Sr/Ba] ratios observed in our sample stars, as seen in Figure~\ref{fig:srba}. But first, we note that about half of our sample stars likely formed from gas enriched by an $r$-process as indicated by their [Sr/Ba] ratios. This would imply that a neutron star merger or another massive stellar $r$-process source was responsible for the observed signatures \citep[e.g.,][]{NSM_Lattimer,NSM_Eichler,Symbalisty1985,collapsars_1,collapsars_2}. The other half appears to be neutron-capture normal stars which may be the result of neutron-capture processes such as an $r$-process yield that was highly diluted in the surrounding gas, a limited $r$-process or an ``incomplete'' $s$-process in massive stars. We discuss the latter two options in more detail below. Both these processes are associated with massive stars (possibly also the $r$-process) that would explode as type\,II SNe on short timescales. This would provide fast enrichment across the early Universe and the original host systems of our metal-poor stars.

\subsubsection{Limited $r$-process in Massive Stars}

We only find two stars to be likely limited $r$-process stars, HE~2214$-$6127 and HE~2332$-$3039. \citet{izutani} computed Sr, Y and Zr yields of the weak/limited $r$-process. In order to reproduce observations of metal-poor stars with excesses in these lighter neutron-capture elements (compared to heavier ones) they find that massive 25\,M$_{\odot}$ energetic hypernovae could be the sites of this process. They explore some of their model parameters and present results for a variation of ejected matter and electron fraction $Y_{e}$. Less massive, normal SNe appear to not produce enough light neutron-capture elements unless an extremely high entropy is assumed for the nucleosynthetic processing. 

We employ the limited $r$-process yields from table~3 presented in \citet{izutani} to calculate what the abundances of a metal-poor star would be assuming that it formed from material enriched by such a process. We calculate [Sr/H] ratios (we deliberately choose not to calculate values with respect to Fe as iron is produced completely independently of the neutron-capture elements). We assume that the Sr yields are mixed into a hydrogen cloud of $10^5$\,M$_{\odot}$ whose size is cosmologically motivated (see also Section~\ref{sec:memps}). 

The range in theoretical Sr yields leads to predictions for observed values of $-4.8<\mbox{[Sr/H]}<-3.9$. Only few stars fall into this range. If we were to assume a slightly smaller gas mass, this range would become higher and could potentially explain the enrichment of HE~2214$-$6127 ([Sr/H] = $-3.01$) and HE~2332$-$3039 ([Sr/H] = $-3.03$), if physically realistic. We note that no yields are provided for heavier neutron-capture elements such as Ba.

\subsubsection{$s$-Process in Massive Rotating Low-Metallicity Stars} 

To possibly explain the bulk of our stars stars (neutron-capture normal as well as stars currently classified as $r$-I; see overlap regions in Figure~\ref{fig:srba}) with their neutron-capture abundance that are neither significantly enhanced nor extraordinarily deficient, we consider the following enrichment scenario for early neutron-capture element production. 
\citet{pignatari} calculated $s$-process nucleosynthesis occurring in massive, rotating low-metallicity (``MRL'') stars. Stellar rotation induces mixing that has been shown to produce large amounts of $^{14}$N in massive low-metallicity objects \citep{meynet06,hirschi07}. A fraction of this nitrogen is mixed in the He-burning core and converted to neon. This process provides free neutrons which are essential for the operation of the $s$-process. The amount of $^{22}$Ne, however, is easily affected by the initial mass of the star, by the rotation speed as well as nuclear reaction rate uncertainties.

The massive non-rotation models of \citet{pignatari} are extremely deficient of neutron-capture elements. But the $s$-process yields (e.g., Sr and Ba) for stars with $\metal <-3.0$ are orders of magnitudes higher in fast rotating 25\,M$_{\odot}$ stars compared to non-rotating stars of the same mass. These drastic abundance changes are primarily dependent of the amount of available primary Ne nuclei. While the Sr production becomes saturated at high rotation speeds irrespective of the metallicity of the model, the Ba yields in the $\metal =-3.0$ model increase by more than 3\,dex from the non-rotating to the rotating cases. For comparison, at solar metallicity, the weak-$s$ process operates in massive stars also but it becomes very inefficient at low metallicity due to decreased amounts of required $^{22}$Ne and Fe seeds (e.g., \citealt{prantzos_weak_s}). The MRL $s$-process yields are listed in Table~\ref{tab:incompl}.

We note that these yields are only those of the (secondary) $s$-process component. It neglects any contributions from the primary explosive component which entails a $p$-process component for which the incomplete $s$-process yields act as seed nuclei as well as any $\alpha$-rich freezeout yields. These other components may add to the overall neutron-capture yields as presented in \citet{pignatari} but likely not in large quantities (possibly a few tenths of a dex). Also, these calculations for a 25\,M$_{\odot}$ are progenitor mass dependent, and for a more massive star the yields would linearly increase with stellar mass (M. Pignatari, priv. comm.).

We utilize the yields of \citet{pignatari}, and again assume a $10^5$\,M$_{\odot}$ dilution mass, to calculate theoretical abundance levels for comparison with our observations. Based on values presented in Table~\ref{tab:incompl}, 
[Sr/H] abundances range from $\mbox{[Sr/H]}<-5.7$ to $-4.6$\,dex for the rotating cases, as well as $\mbox{[Ba/H]}<-5.9$ to $-4.6$\,dex. 
If anything, only CD~$-$24{\textdegree}17504 broadly falls into these ranges, making it not unlikely that its neutron-capture elements were created in one single massive progenitor SN (see Section~\ref{sec:memps}). 

The theoretically predicted MRL [Sr/Ba] ratios are high (largely supersolar), ranging from $-0.17<\mbox{[Sr/Ba]}<1.28$, which agrees with values found for the bulk of our sample stars (see Figure~\ref{fig:srba}). This supports massive objects as progenitors for our observed stars. But again, assuming that early star forming gas masses may have been limited, we can not exclude that [Sr/H] and [Ba/H] may each have been higher locally; the ratio would remain the same. For more context, we note that the [Sr/Ba] values found in the classical main-$s$-process stars\footnote{Those stars are in binary systems and only occur with metallicities of $\metal \gtrsim-2.6$. They are not further considered here.} are much lower, $\mbox{[Sr/Ba]}<-0.6$. So are ratios produced by the main $r$-process, $\mbox{[Sr/Ba]}\sim-0.4$ to $0.4$\,dex as found in strongly $r$-process enriched stars. 

Considering all the observations as well as the theoretical predictions for the MRL $s$-process it appears principally plausible that it was responsible for the chemical enrichment of the bulk of our sample stars, although the absolute yields would have needed to be distributed across a gas mass significantly smaller than 10$^5$ solar masses. Alternatively, we can conclude that principally any other process operating in massive stars would be able to explain the observed abundance ratios. These massive stars were most likely the earliest producers of any neutron-capture elements in the Universe. This would also be consistent with the extensive scatter in Sr abundances observed in metal-poor halo stars at the lowest metallicities (e.g., \citealt{McWilliametal, barklem05, aoki05, francois07,roederer13,Frebel2018}).

\subsection{Mono-Enriched Metal-Poor Stars}\label{sec:memps}

Motivated by the results found in Section~\ref{sec:ncaps_origin}, we further aim to define additional criteria that indicate whether a metal-poor star carries the nucleosynthetic signature of just one previous SN based on its Sr and Ba abundances. \citet{Tilman_mono} suggested a diagnostic tool (see their figure~11) to identify "mono-enriched stars" based on the observed [Mg/C] ratio at a given [Fe/H]. Unfortunately for our sample, carbon abundances could not be measured for all stars, which somewhat limits the use of this tool. 

\begin{figure}
\begin{center}
\includegraphics[clip=true,width=8.5cm, height=12cm]{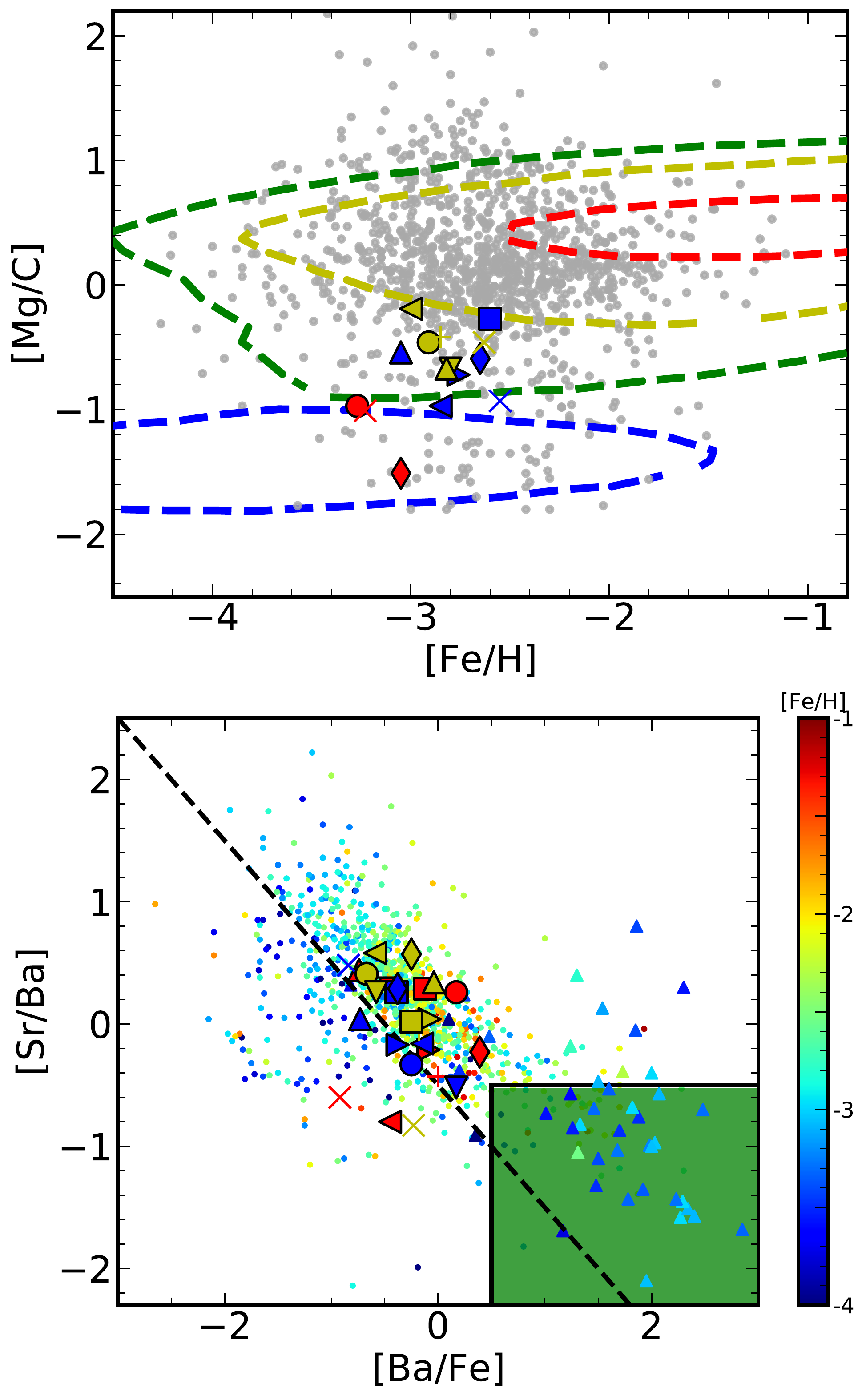} 
\caption{Upper panel shows the observed [Mg/C] ratios as a function of [Fe/H] for literature stars collected by JINAbase with C, Mg, Sr, and Ba measurements. The blue dashed line represents the 3\,$\sigma$ probability of finding mono-enriched stars. The red, yellow, and green dashed lines represent the 1\,$\sigma$, 2\,$\sigma$, and 3\,$\sigma$ of finding multiple-enriched stars, respectively. Bottom panel shows the distributions of the mono- and multiple-enriched stars in the [Ba/Fe] vs. [Sr/Ba] space, color-coded by their [Fe/H] values. Our sample stars are overplotted using same symbols as in Figure~\ref{fig:chemabunds}. A clear separation between the can be mono- and multi-enriched stars can not be drawn in the [Ba/Fe] vs. [Sr/Ba] space. However, the mono-enriched stars seem to populate a shaded region, which is defined by [Ba/Fe] $> 1.0$, [Sr/Ba] $< -0.5$, and [Fe/H] $< -2.7$. Only HE~0232$-$3755 fulfills this criterion, making it a mono-enriched star. Dashed-line denotes [Sr/Ba] = 0.50 -[Ba/Fe] as used in Figure~\ref{fig:chemabunds}.}
\label{fig:mono}
\end{center}
\end{figure}

The upper panel of Figure~\ref{fig:mono} shows [Mg/C] ratios as a function of [Fe/H] for our sample star with theoretically  predicted contours from \citet{Tilman_mono}. The blue dashed line represents the 3\,$\sigma$ probability of a mono-enrichment origin. The red, yellow, and green dashed lines represent the 1\,$\sigma$, 2\,$\sigma$, and 3\,$\sigma$ confidence intervals of the probability contours of multi-enrichment origins, respectively. The contours separate so it is possible to easily identify mono-enriched from multi-enriched stars in this diagram. We overplotted our sample stars using the same symbols in Figure~\ref{fig:chemabunds} and find that four (within error bars) out of the 15 stars (those with available carbon measurements) indeed most likely formed from gas chemically enriched by just one SN event (G64$-$12, CD~$-$24{\textdegree}17504, HE~0232$-$3755, and HE~1245$-$0430). This corresponds to about 30\% of mono-enriched stars in our sample.

We then explored the possibility of the mono-enriched stars having unique [Sr/Ba] ratios that would make their identification more straightforward. We also wanted to check how this would map onto our earlier conclusions regarding single progenitors of our sample stars.

For that, we retrieved all JINAbase metal-poor stars with available C, Mg, Sr, and Ba abundance measurements. This selection yielded a sample of 1099 stars. We overplotted the observed [Mg/C] ratios as a function of [Fe/H] for this literature sample. 44 stars are mono-enriched, as seen in Figure~\ref{fig:mono}. In the bottom panel in Figure~\ref{fig:mono}, we then plot the two mono- and multi-enriched stellar populations in the [Sr/Ba] vs. [Ba/Fe] diagram to investigate if there is a separation between the two populations. Filled triangles and circles represent the mono- and multi-enriched populations, respectively. Symbols are color-coded based on their [Fe/H] values.

The two populations seem to be overlapping in the [Sr/Ba]-[Ba/Fe] diagram. However, a considerable number of mono-enriched metal-poor stars are preferentially located at $\mbox{[Ba/Fe]} > 0.5$, [Sr/Ba] $< -0.5$, and [Fe/H] $< -2.7$ in the Figure (the green shaded area). Unfortunately, non of our sample stars reside within that region. In general, it seems that the [Sr/Ba]-[Ba/Fe] space is not very helpful for making robust conclusion on the number of the SN events responsible for the chemical enrichment of our target stars. Therefore, we hold on our results found using the [Mg/C] vs. [Fe/H] space.

\subsection{Kinematic and orbital properties of the star sample}

Adding orbital information to chemical abundance analysis results can be helpful to further identify the origins of the observed element signatures \citep[e.g.,][]{Mardini2022b,Mardini2022}. Therefore, we investigated the dynamical origins of our sample stars. We retrieved the proper motions ($\mu_{\alpha}\cos\delta$, $\mu_{\delta}$) and parallaxes ($\varpi$) from Gaia EDR3 \citep{Gaia_EDR3}. We adopted the line-of-sight velocities ($V_r$) reported in Table~\ref{Tab:obs}. We corrected $\varpi$ as suggested in \citet{Lindegren_Parallax_2021} and derived heliocentric distances using an exponentially decreasing space density prior as presented in \citet{Mardini2022}. With these quantities we can construct the full space motions of our sample. 

We assume that the Sun is located at R$_\odot =8.178 \pm 0.013$\,kpc from the Galactic center \citep{Gravity_Collaboration2019}, $z_{\odot}= 20.8 \pm 0.3$\,pc above the Galactic plane, and has peculiar motion $U_{\odot} =11.1 \pm 0.72$\,km\,s$^{-1}$ \citep{Bennett2019}, $V_{\odot}= 12.24 \pm 0.47$\,km\,s$^{-1}$, and $W_{\odot}= 7.25 \pm 0.36$\,km\,s$^{-1}$ \citep{Schonrich2010}. We take V$_{LSR}$ = $220$ km\,s$^{-1}$ \citep{Kerr1986}. We used \texttt{The-ORIENT}\footnote{\url{https://github.com/Mohammad-Mardini/The-ORIENT}} \citep[for more details, we refer the readers to ][]{Mardini_2020,Mardini2022b} to integrate the corresponding stellar orbits, apocentric ($\rapo$) and pericentric ($\rperi$) radii, the maximum offset from the Galactic midplane ($\zmax$), and eccentricity, defined as $e = (\rapo - \rperi) / (\rapo + \rperi)$.

\begin{figure*}                        
\begin{center}
\includegraphics[clip=true,width=\textwidth,height=21cm]{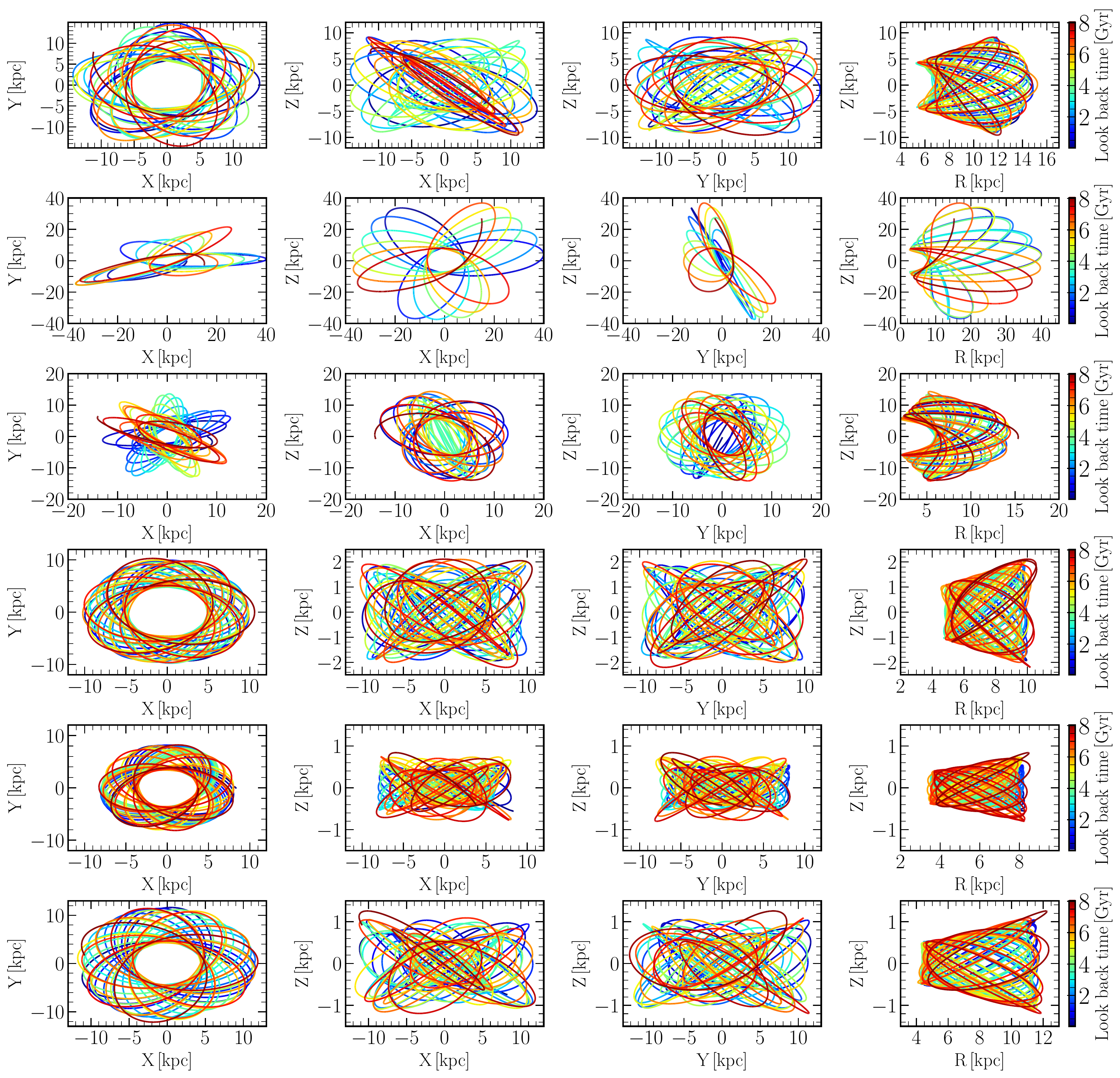}
\caption{Halo star orbital histories for CS~30302-145 (first row), and CD~$-$24{\textdegree}17504 (second row), and HE~2045$-$5057 (third row). Disk star orbital histories for HE~2133$-$0421 (fourth row), HE~0223$-$2814 (fifth row), and CS~29504-018 (sixth row).}
\label{fig:orbits}
\end{center}
\end{figure*}

Employing our action-velocity separation tool developed by \citet{Mardini2022}, we assign individual stars to one of the various Galactic components (thin disk, thick disk, and halo). We find that all of our stars belong to the Galactic halo except CS~29504-018, HE~0223$-$2814, and HE~2133$-$0421 which have disk-like kinematics. Figure~\ref{fig:orbits} shows the orbital histories of several example stars. The outer halo-like nature for CS~30302-145 (first row), CD~$-$24{\textdegree}17504 (second row), and HE~2045$-$5057 (third row) is clearly seen. So is the  disk-like nature of HE~2133$-$0421 (fourth row), HE~0223$-$2814 (fifth row), and CS~29504-018 (sixth row).

For the three disk-like kinematics stars, only one has a C measurement which indicates HE~2133$-$0421 to be a multi-enriched star. This aligns with it residing in the disk now and (likely) having formed in the disk where multiple SNe would have contributed to the enrichment of the birth cloud, with the disk having a high star formation rate. For the 24 halo-like kinematics stars (14 of which have C abundances), four (G64$-$12, CD~$-$24{\textdegree}17504, HE~0232$-$3755, and HE~1245$-$0430) stars are mono-enriched and 10 are multi-enriched.

The orbital histories for the stars (CS~30302-145, CD~$-$24{\textdegree}17504, and HE~2045$-$5057) that reside in the UFDs region (see Figure~\ref{fig:srba}) indicate an outer-halo kinematic. This adds additional line of evidence for the mono-enrichment scenario of CD~$-$24{\textdegree}17504. Also, the extremely low Sr and Ba abundances derived for CD~$-$24{\textdegree}17504 suggests that it could have originated in a satellite galaxy similar to the ones that survived to the present day. In \citet{Andales_SASS}, we further investigated low Sr stars as they are the most likely stars that originated in small dwarf galaxies that got accreted in the earliest phase(s) of the history of the Galaxy. The criterion for a star from a "small accreted stellar system" (SASS), is [Sr/H] $<$ $-4.5$. With [Sr/H] $=$ $-4.75$ and [Ba/H] $=$ $-4.15$, CD~$-$24{\textdegree}17504 fulfills this criterion, making it an ancient SASS star from the early Universe. Therefore, it is most likely that CD~$-$24{\textdegree}17504 is one of the first Pop\,II stars to have formed -- before or during the early formation phase of the Milky Way.

\section{Conclusions and Summary}\label{Sec:summary}
We have presented a chemo-dynamical analysis of twenty seven metal-poor main-sequence stars with [Fe/H] $<-2.5$. These stars were chosen from the program that selects metal-poor candidates \citep{hes4} from the Hamburg/ESO objective-prism survey \citep{hespaperI}. The high-resolution observations were obtained using the MIKE spectrograph on the Magellan-Clay telescope at Las Campanas Observatory, for which twenty stars were followed-up, for the first time.

We derived chemical abundances for 17 elements for our stars. The mean Li abundance ($\langle A(\rm Li) \rangle$ = 2.13) for our sample is in good agreement with values found in other dwarfs at the ``Spite Plateau''. We measured carbon abundances for 15 stars, of which twelve are carbon-enhanced metal-poor stars ([C/Fe] $>0.7$). In general, the light element ($Z<30$) abundances do not differ significantly from those of other similarly metal-poor halo main-sequence stars or giants (see Figure~\ref{fig:chemabunds}), which points to a universal enrichment history of the birth environment(s). Of the heavy elements, we were only able to measure abundances for Sr and Ba which principally agree with those of literature metal-poor stars. 

We further investigated the origins of the neutron-capture elements. We started with the available metal-poor stars data collected by JINAbase, UFDs stars, and abundance criteria from \citet{Frebel2018} for identifying different heavy element nucleosynthesis signatures. The observed [Sr/Ba] ratios for the vast majority of our sample are located in regions in the [Sr/Ba] vs [Ba/Fe] space typically occupied by $r$-I and neutron-capture normal stars, as depicted in Figure~\ref{fig:srba}. For example, two stars, HE~2214$-$6127 and HE~2332$-$3039, have [Sr/H] values roughly in agreement with what is predicted by the limited $r$-process yields presented in \citet{izutani}. The predicted yields of the ``incomplete'' $s$-process presented in \citet{pignatari} broadly matches several stars, but primarily in their [Sr/Ba] ratios. The overall low [Sr/H] and [Ba/H] values are not inconsistent with those of CD~$-$24{\textdegree}17504 but not ideal match exists when assuming a gas mixing mass of 10$^{5}$\,M$_{\odot}$.

Therefore, we conclude that, most likely, the bulk of our stars formed from early gas clouds enriched by either an $r$-process event (e.g., neutron star merger), a limited $r$-process, or an ``incomplete'' $s$-process in massive stars. All these processes have the potential to leave behind signatures that results in observed abundance patterns associated with $r$-I and neutron-capture normal stars. It remains to be seen whether fast enrichment by SNe dominated these enrichment levels among the original host systems of our stars, or whether later time enrichment by neutron star mergers occurred in some ways before our metal-poor stars formed. Finally, HE~0232$-$3755 is most likely an $r$-II star as its [Sr/Ba] ratio can be explained by an $r$-process event. 

We also performed a full orbital history analysis of our sample stars, using astrometric data from Gaia EDR3 \citep{Gaia_EDR3}. This analysis suggests halo-like kinematics for the vast majority of our sample. This points to an accretion origin of these stars. Finally, we used [Mg/C] vs. [Fe/H] criterion by \citet{Tilman_mono} to identify mono-enriched stars in our sample, i.e. stars with just one progenitor. Based on the observed ratios, about 30\% of our sample stars with carbon measurements appear to be mono-enriched and likely formed from gas clouds enriched by just one previous SN. This contrasts the kinematic signature of HE~2133$-$0421, HE~0223$-$2814, and CS~29504-018 which are clearly disk-like in nature. 

Then there are are several interesting stars to mention. Three stars, CS~30302-145, HE~2045$-$5057, and CD~$-$24{\textdegree}17504, have [Sr/Ba] and [Ba/Fe] ratios consistent with those observed in typical UFDs stars. They also all have halo-like kinematics. This strongly suggests that they originated in long-disrupted early dwarf galaxies. Of particular interest in this context is CD~$-$24{\textdegree}17504 which has the lowest Sr and Ba abundances in our sample. Its distinct chemodynamical nature is consistent with an origin in a small accreted stellar system (SASS) that was absorbed by the proto-Milky Way, as laid out in \citet{Andales_SASS}. This suggest that CD~$-$24{\textdegree}17504 is likely among the earliest, and hence oldest, Pop\,II stars to have formed in the Universe. More stars with low Sr and Ba abundance measurements will help to identify more of these ancient survivors from the early universe.

\section*{Acknowledgements}
This work is supported by Basic Research Grant (Super AI) of Institute for AI and Beyond of the University of Tokyo. M.k.M. acknowledges partial support from NSF grant OISE 1927130 (International Research Network for Nuclear Astrophysics/IReNA). A.F. acknowledges support from  NSF CAREER grant AST-1255160 and NSF grant AST-1716251.

This work made use of the NASA's Astrophysics Data System Bibliographic Services.

This work has made use of data from the European Space Agency (ESA) mission
{\it Gaia} (\url{https://www.cosmos.esa.int/gaia}), processed by the {\it Gaia}
Data Processing and Analysis Consortium (DPAC,
\url{https://www.cosmos.esa.int/web/gaia/dpac/consortium}). Funding for the DPAC
has been provided by national institutions, in particular the institutions
participating in the {\it Gaia} Multilateral Agreement.

This research has made use of the SIMBAD database,
operated at CDS, Strasbourg, France.

\section*{Data Availability}
The data underlying this article will be shared on reasonable request to the corresponding author.



\bibliographystyle{mnras}
\bibliography{references}

\begin{table*}
\centering
\caption{Observing Details}
\label{Tab:obs}
\resizebox{\textwidth}{!}{
\begin{tabular}{lrrrrrrrcc}
\toprule
Star& $\alpha$ & $\delta$ & $B$&UT& $t_{\rm {exp}}$&Slit&$v_{\rm{rad}}$&S/N at&S/N at \\
    & (J2000) & (J2000) & mag&date$^a$& sec&size&km\,s$^{-1}$&4070\,{\AA}$^b$&5170\,{\AA}$^b$ \\
\midrule
CD~$-$24{\textdegree}17504  & 23 07 20.2 & $-$23 52 36 &  12.5 & Jul 28, 2009 &  450 & 0.7  &    136.4 &   48  &  64  \\
CS22188$-$033  & 00 51 25.9 & $-$38 12 18 &  13.6 & Feb 05, 2009 &  900 & 0.7  &     18.3 &   15  &  30  \\
CS29504$-$018  & 01 32 54.8 & $-$32 55 25 &  14.1 & Jul 27, 2009 &  800 & 0.7  &   $-$4.8 &   37  &  64  \\
               &            &             &       & Jul 28, 2009 & 1200 & 0.7  &   $-$5.0 &       &      \\
CS29519$-$133  & 02 19 22.8 & $-$48 04 38 &  13.2 & Jul 27, 2009 &  600 & 0.7  &    123.5 &   68  &  74  \\
CS30302$-$145  & 19 40 52.2 & $-$48 39 19 &  14.8 & Jul 27, 2009 & 1500 & 0.7  &    196.8 &   73  &  93  \\
               &            &             &       & Jul 28, 2009 & 2100 & 0.7  &    196.4 &       &      \\
G64$-$12       & 13 40 02.5 & $-$00 02 19 &  11.9 & Feb 06, 2009 &  400 & 0.7  &    443.4 &   84  & 137  \\
HE~0219$-$2056 & 02 21 34.3 & $-$20 42 57 &  15.7 & Feb 05, 2009 & 3600 & 0.7  &  $-$46.5 &   44  &  70  \\
               &            &             &       & Feb 06, 2009 & 2700 & 0.7  &  $-$46.4 &       &      \\
               &            &             &       & Feb 07, 2009 & 2700 & 0.7  &  $-$46.2 &       &      \\
HE~0223$-$2814 & 02 25 17.0 & $-$28 00 46 &  13.0 & Feb 07, 2009 &  600 & 0.7  &    149.5 &   51  &  70  \\
HE~0232$-$3755 & 02 34 16.4 & $-$37 42 18 &  13.6 & Feb 07, 2009 &  900 & 0.7  & $-$238.1 &   90  & 154  \\
HE~0406$-$3120 & 04 08 31.9 & $-$31 12 58 &  14.1 & Feb 07, 2009 & 1200 & 0.7  &  $-$37.8 &   66  & 120  \\
               &            &             &       & Mar 11, 2011 & 1800 & 0.7  &  $-$32.2 &       &      \\
HE~0444$-$2938 & 04 46 19.6 & $-$29 33 06 &  15.3 & Feb 05, 2009 & 3600 & 0.7  &    247.2 &   32  &  55  \\
HE~0526$-$4059 & 05 28 23.0 & $-$40 56 50 &  14.7 & Mar 23, 2010 & 3600 & 1.0  &    177.4 &   51  &  77  \\
HE~1200$-$0009 & 12 02 53.6 & $-$00 26 20 &  16.2 & Feb 05, 2009 & 3600 & 0.7  &    314.8 &   58  & 115  \\
               &            &             &       & Feb 06, 2009 & 5400 & 0.7  &    314.6 &       &      \\
               &            &             &       & Feb 07, 2009 & 3600 & 0.7  &    314.1 &       &      \\
HE~1214$-$1050 & 12 16 44.9 & $-$11 06 46 &  15.5 & Feb 06, 2009 & 3600 & 0.7  &    128.9 &   42  &  61  \\
               &            &             &       & Mar 10, 2011 & 3000 & 0.7  &    100.6 &       &      \\
HE~1245$-$0430 & 12 47 51.1 & $-$04 46 38 &  15.5 & Feb 06, 2009 & 3600 & 0.7  & $-$189.0 &   45  &  60  \\
HE~1309$-$1113 & 13 11 52.1 & $-$11 29 21 &  15.3 & Feb 20, 2009 & 2260 & 0.7  &  $-$43.0 &   35  &  52  \\
HE~1401$-$0010 & 14 04 03.4 & $-$00 24 25 &  13.9 & Aug 07, 2010 & 1500 & 1.0  &    387.4 &   54  &  75  \\
HE~1436$-$0654 & 14 39 36.8 & $-$07 07 47 &  15.2 & Jul 28, 2009 & 2100 & 0.7  &  $-$36.2 &   51  &  57  \\
               &            &             &       & Mar 13, 2011 & 1800 & 0.7  &  $-$90.4 &       &      \\
HE~1929$-$6715 & 19 34 58.3 & $-$67 09 22 &  14.4 & Jul 28, 2009 &  900 & 0.7  &    159.0 &   75  & 106  \\
               &            &             &       & Aug 05, 2010 & 1200 & 0.7  &    158.1 &       &      \\
HE~2045$-$5057 & 20 48 46.7 & $-$50 46 29 &  15.1 & Aug 05, 2010 & 1800 & 0.7  & $-$102.3 &   77  & 137  \\
               &            &             &       & Aug 06, 2010 & 1200 & 0.7  & $-$102.3 &       &      \\
HE~2130$-$4852 & 21 33 58.9 & $-$48 39 28 &  15.3 & Jul 27, 2009 & 2000 & 0.7  & $-$195.1 &   86  & 115  \\
               &            &             &       & Aug 05, 2010 & 2700 & 0.7  & $-$201.8 &       &      \\
HE~2133$-$0421 & 21 36 15.1 & $-$04 08 17 &  15.9 & Aug 06, 2010 & 3300 & 0.7  &   $-$2.8 &   45  &  86  \\
HE~2141$-$2916 & 21 44 36.8 & $-$29 02 40 &  15.1 & Jul 27, 2009 & 2000 & 0.7  &     65.0 &   39  &  71  \\
HE~2214$-$6127 & 22 18 16.0 & $-$61 12 17 &  15.8 & Aug 08, 2010 & 3600 & 1.0  &  $-$29.3 &   45  &  57  \\
HE~2231$-$0635 & 22 34 07.6 & $-$06 20 00 &  13.6 & Jul 28, 2009 & 1200 & 0.7  & $-$292.1 &   26  &  34  \\
HE~2308$-$3543 & 23 11 32.3 & $-$35 26 42 &  15.8 & Jul 27, 2009 & 2400 & 0.7  & $-$123.9 &   95  &  134 \\
               &            &             &       & Jul 28, 2009 & 3300 & 0.7  & $-$124.3 &       &      \\
               &            &             &       & Aug 05, 2010 & 1800 & 0.7  & $-$124.4 &       &      \\
HE~2332$-$3039 & 23 35 06.9 & $-$30 22 54 &  15.6 & Aug 06, 2010 & 2100 & 0.7  &  $-$12.6 &   74  &  89   \\
\bottomrule
\multicolumn{3}{l}{$^a$ At the start of the observation.}\\
\multicolumn{3}{l}{\footnotesize$^b$ $S/N$ is given per pixel.}\\
\end{tabular}
}
\end{table*}

\begin{table}
\caption{Equivalent widths measurements}
\label{Tab:Eqw}
\resizebox{\textwidth}{!}{
\begin{tabular}{lrrrrrrrcc}
\toprule
El.& $\lambda$ & $\chi$ & $\log gf$ & EW &$\log\epsilon$                 &EW &$\log\epsilon$                     & EW &$\log\epsilon$ \\
\cmidrule(rl){5-6} \cmidrule(rl){7-8} \cmidrule(rl){9-10}
   &           &        &           & \multicolumn{2}{c}{HE~0219$-$2056} &  \multicolumn{2}{c}{HE~1200$-$0009}   &  \multicolumn{2}{c}{G64$-$12} \\
\midrule  
Na\,I &  5889.95 &  0.00 &    0.101 &   34.8 &      2.80   &  41.1 &  2.97    &  38.2 &   2.97 \\
Na\,I &  5895.92 &\ldots& \ldots  &\ldots & \ldots     &  26.9 &  3.02    &  15.2 &   2.39 \\
Mg\,I &  4703.00 &  4.34 & $-$0.670 &\ldots & \ldots     &\ldots&\ldots   &   8.6 &   4.91 \\
Mg\,I &  5172.70 &  2.71 & $-$0.380 &   78.5 &      4.48   &  73.1 &  4.47    &  78.9 &   4.68 \\
Mg\,I &  5183.62 &  2.72 & $-$0.160 &   91.4 &      4.48   &  84.5 &  4.43    &  91.7 &   4.69 \\
Mg\,I &  5528.40 &  4.35 & $-$0.490 &    8.5 &      4.64   &\ldots&\ldots   &\ldots&\ldots \\
Al\,I &  3961.52 &  0.00 & $-$0.340 &   45.5 &      2.75   &  15.5 &  2.20    &  23.1 &   2.48 \\
Si\,I &  3905.53 &  1.91 & $-$1.090 &   79.6 &      4.63   &  38.9 &  4.05    &  59.8 &   4.50 \\
Ca\,I &  4226.74 &  0.00 &    0.240 &   86.6 &      3.18   &  82.1 &  3.19    &  85.7 &   3.46 \\
Ca\,I &  4302.54 &  1.90 &    0.280 &    8.5 &      3.16   &\ldots&\ldots   &  18.1 &   3.65 \\
Ca\,I &  4454.79 &  1.90 &    0.260 &   12.1 &      3.34   &\ldots&\ldots   &  11.7 &   3.44 \\
Ca\,I &  5588.76 &  2.53 &    0.358 &    5.8 &      3.42   &\ldots&\ldots   &\ldots&\ldots \\
Ca\,I &  6122.22 &  1.89 & $-$0.320 &    4.7 &      3.41   &\ldots&\ldots   &\ldots&\ldots \\ 
Sc\,II&  4246.82 &  0.32 &    0.242 &   18.3 &   $-$0.09   &  17.1 &  0.10    &  17.3 &   0.12 \\
Ti\,II&  3759.30 &  0.61 &    0.270 &\ldots & \ldots     &  56.9 &  1.59    &  67.7 &   1.92 \\
Ti\,II&  3761.33 &  0.57 &    0.170 &   76.4 &      1.81   &  47.6 &  1.49    &  65.5 &   1.94 \\
Ti\,II&  3900.54 &  1.13 & $-$0.450 &   29.2 &      2.02   &\ldots&\ldots   &\ldots&\ldots \\
Ti\,II&  4012.39 &  0.57 & $-$1.610 &    8.5 &      2.01   &\ldots&\ldots   &\ldots&\ldots \\
Ti\,II&  4300.05 &  1.18 & $-$0.490 &   18.1 &      1.80   &\ldots&\ldots   &  18.3 &   2.03 \\
Ti\,II&  4395.03 &  1.08 & $-$0.510 &   29.7 &      2.00   &  12.7 &  1.76    &  23.2 &   2.09 \\
Ti\,II&  4399.77 &  1.24 & $-$1.290 &\ldots & \ldots     &\ldots&\ldots   &   5.2 &   2.26 \\
Ti\,II&  4417.72 &  1.16 & $-$1.160 &\ldots & \ldots     &\ldots&\ldots   &   8.2 &   2.27 \\
Ti\,II&  4443.81 &  1.08 & $-$0.700 &   25.3 &      2.09   &  10.3 &  1.85    &  13.8 &   1.99 \\
Ti\,II&  4468.51 &  1.13 & $-$0.600 &   16.1 &      1.79   &\ldots&\ldots   &  18.4 &   2.08 \\
Ti\,II&  4501.28 &  1.12 & $-$0.760 &   24.0 &      2.15   &   8.6 &  1.85    &  14.4 &   2.10 \\
Ti\,II&  4533.97 &  1.24 & $-$0.640 &   13.4 &      1.84   &  14.1 &  2.08    &  14.0 &   2.08 \\
Ti\,II&  4563.77 &  1.22 & $-$0.820 &   10.7 &      1.88   &\ldots&\ldots   &  11.8 &   2.15 \\
Ti\,II&  4571.98 &  1.57 & $-$0.340 &   28.7 &      2.25   &\ldots&\ldots   &  12.7 &   2.02 \\ 
Cr\,I &  4254.33 &  0.00 & $-$0.110 &   26.2 &      2.34   &  10.6 &  1.97    &  15.9 &   2.25 \\
Cr\,I &  4274.79 &  0.00 & $-$0.230 &\ldots & \ldots     &  10.8 &  2.09    &  18.0 &   2.43 \\
Cr\,I &  4289.72 &  0.00 & $-$0.361 &   11.8 &      2.17   &   7.9 &  2.08    &   9.5 &   2.23 \\
Mn\,I &  4030.75 &  0.00 &  HFS     &   syn  &   $<$1.80   &   syn &$<$1.6    &   syn &   1.60 \\
\bottomrule
\end{tabular}
}\\
This table is published in its entirety in the electronic edition of the paper. A portion is shown here for guidance regarding its form and content.
\end{table}

\begin{table*}
\caption{Stellar Parameters}
\label{Tab:stellpar}
\begin{tabular}{l|cccc|cccccc}
\toprule
Star& $T_{\rm{eff,ini}}$ & $\log (g)_{\rm{ini}}$ & $\mbox{[Fe/H]}_{\rm{ini}}$ & $v_{\rm{micr, ini}}$ &$T_{\rm{eff,corr}}$&$\log (g)_{\rm{corr}}$ &$\mbox{[Fe/H]}_{\rm{corr}}$& $v_{\rm{micr,corr}}$ &$T_{\rm{eff,BP-G}}$&$\log (g)_{\rm{Gaia\_plx}}$\\
&k&dex&dex&[km\,s$^{-1}$] &k&dex&dex&[km\,s$^{-1}$]&k&dex \\
\midrule 
CS22188$-$033  & 6200 & 4.40 & $-2.65$ & 1.45 & 6250 & 4.50 & $-2.60$ & 1.45 & 6265 & 4.45 \\ 
CS29504$-$018  & 6240 & 3.70 & $-2.75$ & 1.75 & 6286 & 3.75 & $-2.70$ & 1.80 & 6387 & 4.01 \\
CS29519$-$133  & 6050 & 3.50 & $-2.70$ & 1.45 & 6115 & 3.75 & $-2.65$ & 1.40 & 6187 & 4.07 \\
CS30302$-$145  & 6220 & 3.60 & $-2.95$ & 1.70 & 6268 & 3.75 & $-2.95$ & 1.70 & 6390 & 3.90 \\ 
HE~0219$-$2056 & 6569 & 4.15 & $-2.95$ & 1.55 & 6582 & 4.20 & $-3.16$ & 1.55 & 6487 & 3.99 \\
HE~0223$-$2814 & 6180 & 3.45 & $-2.55$ & 1.65 & 6232 & 3.70 & $-2.50$ & 1.65 & 6228 & 3.95 \\
HE~0232$-$3755 & 6140 & 4.35 & $-3.05$ & 1.45 & 6196 & 4.40 & $-3.05$ & 1.50 & 6292 & 4.46 \\
HE~0406$-$3120 & 6490 & 4.20 & $-2.65$ & 1.40 & 6511 & 4.20 & $-2.65$ & 1.40 & 6373 & 3.95 \\
HE~0444$-$2938 & 6390 & 4.05 & $-2.60$ & 1.75 & 6421 & 4.00 & $-2.55$ & 1.80 &\ldots&\ldots\\
HE~0526$-$4059 & 6150 & 3.70 & $-3.15$ & 1.50 & 6205 & 3.75 & $-3.38$ & 1.55 & 6319 & 3.95 \\
HE~1200$-$0009 & 6430 & 4.00 & $-3.46$ & 1.65 & 6457 & 4.05 & $-3.42$ & 1.65 & 6338 & 4.13 \\
HE~1214$-$1050 & 5900 & 3.80 & $-2.95$ & 1.35 & 5980 & 3.80 & $-3.05$ & 1.40 & 6080 & 3.95 \\
HE~1245$-$0430 & 6250 & 3.50 & $-2.85$ & 1.35 & 6295 & 3.80 & $-2.85$ & 1.35 & 6466 & 4.07 \\
HE~1309$-$1113 & 6350 & 3.85 & $-2.50$ & 1.55 & 6385 & 3.90 & $-2.76$ & 1.55 & 6431 & 4.04 \\
HE~1401$-$0010 & 6160 & 3.35 & $-2.60$ & 1.75 & 6214 & 3.75 & $-2.60$ & 1.70 & 6245 & 4.00 \\
HE~1436$-$0654 & 6140 & 3.55 & $-2.70$ & 1.70 & 6196 & 3.85 & $-2.65$ & 1.70 & 6336 & 4.02 \\
HE~1929$-$6715 & 6050 & 3.60 & $-2.70$ & 1.35 & 6115 & 3.65 & $-2.91$ & 1.35 & 6236 & 3.75 \\
HE~2045$-$5057 & 6145 & 3.75 & $-2.40$ & 1.30 & 6200 & 3.85 & $-2.63$ & 1.35 & 6266 & 3.90 \\
HE~2130$-$4852 & 5760 & 3.50 & $-2.90$ & 1.45 & 5854 & 3.65 & $-2.85$ & 1.50 & 5833 & 3.57 \\
HE~2133$-$0421 & 6000 & 3.70 & $-2.85$ & 1.45 & 6070 & 3.70 & $-2.80$ & 1.50 & 6106 & 3.82 \\
HE~2141$-$2916 & 5940 & 3.50 & $-2.88$ & 1.50 & 6016 & 3.60 & $-2.82$ & 1.55 & 6096 & 3.65 \\
HE~2214$-$6127 & 5540 & 3.15 & $-3.10$ & 1.60 & 5656 & 3.55 & $-3.00$ & 1.60 & 5726 & 3.53 \\
HE~2231$-$0635 & 6350 & 3.90 & $-2.31$ & 1.35 & 6385 & 3.95 & $-2.26$ & 1.35 & 6391 & 4.07 \\
HE~2308$-$3543 & 5930 & 4.45 & $-3.66$ & 1.30 & 6007 & 4.60 & $-3.61$ & 1.30 & 6091 & 4.48 \\
HE~2332$-$3039 & 5940 & 3.85 & $-3.43$ & 1.45 & 6016 & 3.85 & $-3.35$ & 1.45 & 6102 & 3.85 \\
CD~$-$24{\textdegree}17504  & 6210 & 3.60 & $-3.26$ & 1.35 & 6259 & 3.80 & $-3.23$ & 1.35 & 6370 & 3.90 \\
G64$-$12       & 6440 & 4.25 & $-3.32$ & 1.30 & 6466 & 4.25 & $-3.27$ & 1.35 & 6411 & 4.21 \\
\bottomrule
\end{tabular}
\end{table*}

\begin{table*}
\caption{Magellan/MIKE abundances}
\label{Tab:abundances}
\begin{tabular}{lrrrrrrrrrrrrrrrrrrrrrrrrrrr}
\toprule
&\multicolumn{5}{c}{HE~0219$-$2056}&& \multicolumn{5}{c}{HE~1200$-$0009}&&\multicolumn{5}{c}{G64$-$12}\\
\cline{2-6} \cline{8-12}\cline{14-18}
Elem.&$\lg\epsilon(\mbox{X})$&[X/H]&[X/Fe]&$\sigma$&$N$&&$\lg\epsilon (\mbox{X})$&[X/H]&[X/Fe]&$\sigma$&$N$&&$\lg\epsilon (\mbox{X})$&[X/H]&[X/Fe]&$\sigma$&$N$\\
\midrule 
Li\,I &   2.14 &\ldots  & \ldots &0.15   &  1   &&   2.16 &\ldots  & \ldots & 0.15 &   1  &&   2.19&\ldots  &\ldots  & 0.15& 1\\
C (CH)&   7.02 &$-$1.41 &$<$1.75 & \ldots&  1   &&   6.41 &$-$2.02 &$<$1.40 &\ldots&   1  &&   6.54&$-$1.89 &   1.38 & 0.10& 1\\
Na\,I &   2.68 &$-$3.56 &$-$0.40 &0.20   &  2   &&   2.94 &$-$3.30 &   0.12 &0.07  &   2  &&   2.97&$-$3.27 &   0.00 & 0.10& 2 \\
Mg\,I &   4.86 &$-$2.74 &   0.42 &0.12   &  4   &&   4.37 &$-$3.23 &   0.19 &0.10  &   6  &&   4.74&$-$2.86 &   0.41 & 0.12& 7 \\
Al\,I &   2.62 &$-$3.83 &$-$0.67 &0.10   &  1   &&   2.32 &$-$4.13 &$-$0.71 &0.05  &   3  &&   2.47&$-$3.98 &$-$0.71 & 0.10& 1 \\
Si\,I &   4.30 &$-$3.21 &$-$0.05 &0.10   &  1   &&   4.08 &$-$3.43 &$-$0.01 &0.10  &   1  &&   4.42&$-$3.09 &   0.18 & 0.10& 1 \\
Ca\,I &   3.67 &$-$2.67 &   0.49 &0.13   &  3   &&   3.55 &$-$2.79 &   0.63 &0.19  &   8  &&   3.56&$-$2.78 &   0.49 & 0.12& 12 \\
Sc\,II&   0.57 &$-$2.58 &   0.58 &0.15   &  1   &&$-$0.56 &$-$3.71 &$-$0.29 &0.15  &   1  &&   0.12&$-$3.03 &   0.24 & 0.15& 2 \\
Ti\,I &   2.17 &$-$2.78 &   0.38 &0.10   &  1   &&   1.97 &$-$2.64 &   0.47 &0.21  &   2  &&   2.34&$-$2.61 &   0.66 & 0.11& 3\\
Ti\,II&   2.32 &$-$2.63 &   0.53 &0.09   & 11   &&   2.14 &$-$2.81 &   0.61 &0.12  &  16  &&   2.12&$-$2.83 &   0.44 & 0.11& 22\\
Cr\,I &   2.39 &$-$3.25 &$-$0.09 &0.11   &  3   &&   3.64 &$-$3.60 &$-$0.18 &0.07  &   3  &&   2.14&$-$3.50 &$-$0.23 & 0.13& 6 \\
Mn\,I & \ldots &\ldots  &\ldots  &\ldots &\ldots&& \ldots &\ldots  &\ldots  &\ldots&\ldots&&   1.54&$-$3.89 &$-$0.62 & 0.05& 3 \\
Fe\,I &   4.34 &$-$3.16 &\ldots  &0.14   & 44   &&   4.08 &$-$3.42 & \ldots &0.11  &  35  &&   4.23&$-$3.27 & \ldots & 0.05&76 \\
Fe\,II&   4.35 &$-$3.15 &$-$0.01 &0.06   &  2   &&   4.06 &$-$3.44 &$-$0.02 &0.08  &   2  &&   4.22&$-$3.28 &$-$0.01 & 0.05& 5 \\
Co\,I & \ldots &\ldots  &\ldots  &\ldots &\ldots&&   0.93 &$-$3.06 &   0.36 &0.10  &   1  &&   2.43&$-$2.56 &   0.71 & 0.10& 1 \\
Ni\,II&   3.35 &$-$2.87 &   0.29 &0.06   &  2   &&   3.02 &$-$3.20 &   0.22 &0.11  &   6  &&   3.04&$-$3.18 &   0.09 & 0.10&10 \\
Sr\,II&$-$0.58 &$-$3.45 &$-$0.30 &0.05   &  1   &&$-$0.90 &$-$3.77 &$-$0.35 &0.13  &   1  &&   0.03&$-$2.84 &   0.43 & 0.06& 1 \\
Ba\,II&$-1.06$ &$-$3.24 &$-$0.09 &0.16   &  1   &&$-$1.07 &$-$3.25 &   0.17 &0.05  &   1  &&$-$0.92&$-$3.10 &   0.17 & 0.05& 1 \\ 
\bottomrule
\end{tabular}
\\
This table is published in its entirety in the electronic edition of the paper. A portion is shown here for guidance regarding its form and content.
\end{table*}

\begin{table}
\caption{Example Abundance Uncertainties}
\label{Tab:err}
\begin{tabular}{lrrrrr}
\toprule
Elem.&Random& $\Delta$\mbox{T$_{\rm eff}$}&$\Delta\log g$ &$\Delta v_{micr}$ &  Total\\
     &error & +150\,K                     &$+$0.5\,dex    &+0.3\,km\,s$^{-1}$& $\sigma$\\
\midrule 
\multicolumn {5}{c}{G64$-$12} \\ \hline
O \textsc{I} & & & & & \\
Na \textsc{I} & & $-0.10$ & $-0.01$ & $-0.02$ & 0.10 \\ 
Mg \textsc{I} & & $-0.08$ & $-0.02$ & $-0.02$ & 0.08 \\ 
Al \textsc{I} & &$-0.11$ & 0.00 & $-0.01$ & 0.11 \\ 
Si \textsc{I} & &$-0.11$ & $-0.01$ & $-0.04$ & 0.12 \\ 
Ca \textsc{I} & & $-0.08$ & $-0.01$ & $-0.01$ & 0.08 \\ 
Sc \textsc{II} & & $-0.07$ & 0.09 & 0.00 & 0.11 \\ 
Ti \textsc{I} & & $-0.13$ & 0.00 & $-0.01$ & 0.13 \\ 
Ti \textsc{II} & & $-0.08$ & 0.09 & $-0.04$ & 0.13 \\ 
Cr \textsc{I} & & $-0.13$ & 0.00 & 0.00 & 0.13 \\ 
Cr \textsc{II} & & $-0.06$ & 0.08 & $-0.01$ & 0.10 \\ 
Mn \textsc{I} & & $-0.14$ & 0.01 & 0.00 & 0.14 \\ 
Fe \textsc{I} & & $-0.12$ & $-0.01$ & $-0.03$ & 0.12 \\ 
Fe \textsc{II} & & $-0.02$ & 0.00 & $-0.01$ & 0.02 \\ 
Co \textsc{I} & & $-0.13$ & 0.01 & $-0.01$ & 0.13 \\ 
Ni \textsc{I} & & $-0.17$ & $-0.02$ & $-0.06$ & 0.18 \\ 
Sr \textsc{II} & & $-0.09$ & 0.08 & $-0.08$ & 0.14 \\ 
Ba \textsc{II} & & $-0.10$ & 0.09 & 0.00 & 0.13 \\\hline
\multicolumn {5}{c}{CS29519$-$133}\\ \hline
Na \textsc{I} & & 0.10 & $-0.02$ & $-0.03$ & 0.11 \\ 
Mg \textsc{I} & & 0.09 & $-0.07$ & $-0.03$ & 0.12 \\ 
Al \textsc{I} & & 0.12 & 0.00 & $-0.01$ & 0.12 \\ 
Si \textsc{I} & & 0.13 & $-0.06$ & $-0.08$ & 0.16 \\ 
K \textsc{I} & & 0.09 & 0.00 & 0.00 & 0.09 \\ 
Ca \textsc{I} & & 0.08 & $-0.02$ & $-0.02$ & 0.08 \\ 
Ca \textsc{II} & & 0.00 & 0.00 & 0.00 & 0.00 \\ 
Sc \textsc{II} & & 0.08 & 0.10 & $-0.01$ & 0.13 \\ 
Ti \textsc{I} & & 0.14 & 0.00 & $-0.01$ & 0.14 \\ 
Ti \textsc{II} & & 0.08 & 0.08 & $-0.05$ & 0.12 \\ 
V \textsc{I} & & 0.10 & 0.00 & 0.00 & 0.10 \\ 
Cr \textsc{I} & & 0.15 & $-0.02$ & $-0.04$ & 0.16 \\ 
Mn \textsc{I} & & 0.16 & 0.00 & 0.00 & 0.16 \\ 
Fe \textsc{I} & & 0.13 & $-0.02$ & $-0.05$ & 0.14 \\ 
Fe \textsc{II} & & 0.03 & 0.10 & $-0.01$ & 0.10 \\ 
Co \textsc{I} & & 0.15 & 0.00 & 0.00 & 0.15 \\ 
Ni \textsc{I} & & 0.19 & $-0.05$ & $-0.07$ & 0.21 \\ 
Sr \textsc{II} & & 0.08 & 0.08 & $-0.09$ & 0.14 \\ 
Ba \textsc{II} & & 0.11 & 0.09 & 0.00 & 0.14 \\
\bottomrule
\end{tabular}
\end{table}

\begin{table*}
\caption{MRL s-Process Predictions for Metal-Poor Stars}
\label{tab:incompl}
\begin{tabular}{lrcccccc}
\toprule
Ratio& Mean atomic & $\mbox{[Fe/H]}=-3.0$ &$\mbox{[Fe/H]}=-4.0$ & $\mbox{[Fe/H]}=-3.0$ &$\mbox{[Fe/H]}=-4.0$ & $\mbox{[Fe/H]}=-3.0$ &$\mbox{[Fe/H]}=-4.0$ \\
& number& non-rotating & non-rotating& slow rotating & slow rotating &fast rotating & fast rotating\\
\midrule
$\mbox{[Sr/H]}$& 87.7  &$-$6.47 &$-$8.68 & $-$4.57 & $-$5.62 & $-$4.70 &$-$5.72\\
$\mbox{[Y/H]}$ & 89.0  &$-$6.89 &$-$8.92 & $-$4.49 & $-$5.40 & $-$4.43 &$-$5.45\\
$\mbox{[Zr/H]}$& 91.3  &$-$7.35 &$-$9.29 & $-$4.66 & $-$5.44 & $-$4.41 &$-$5.40\\
$\mbox{[Mo/H]}$& 96.0  &$-$7.87 &$-$9.62 & $-$4.88 & $-$5.42 & $-$4.60 &$-$5.26\\
$\mbox{[Ba/H]}$&137.4  &$-$7.97 &$-$9.09 & $-$5.85 & $-$5.83 & $-$4.61 &$-$5.55\\
 \hline
[Sr/Ba]        &       & 1.5    &   0.41 &    1.28 &  0.21 & $-$0.09 &$-$0.17\\
\bottomrule
\multicolumn{8}{l}{$^a$ Based on $s$-process yields in rotating massive
low-metallicity stars presented in \citet{pignatari}. They indirectly
give the rotation}\\ 
\multicolumn{8}{l}{speed by the amount of $^{22}$Ne produced through
rotation-induced $^{14}$N and available for the $s$-process. All values
are calculated by}\\ 
\multicolumn{8}{l}{assuming a dilution mass of 10$^{5}$\,M$_{\odot}$, a
C-shell mass of 3\,M$_{\odot}$, and solar abundance from
\citet{asplund09}.}\\
\end{tabular}
\end{table*}

\bsp	
\label{lastpage}
\end{document}